\documentclass[a4paper,11pt,reqno]{article}

\usepackage[margin=2.5cm]{geometry}
\usepackage[svgnames]{xcolor}
\usepackage{tikz}
\usepackage{color}
\usepackage{amssymb}
\usepackage{float}
\usepackage{enumitem}
\usepackage{array}
\usepackage{longtable}
\usepackage{ctable} 
\usepackage{multirow}
\usepackage{mathrsfs}
\usepackage{caption}
\usepackage{footmisc}
\usepackage{stmaryrd}
\usepackage{arydshln}
\usepackage{amsthm,amssymb,amsmath}
\usepackage{enumitem}

\usepackage[colorlinks = true,
linkcolor = blue,
urlcolor  = blue,
citecolor = blue,
anchorcolor = blue]{hyperref} 

\newcolumntype{C}[1]{>{\centering\arraybackslash}m{#1}}
\newcolumntype{L}[1]{>{\arraybackslash}m{#1}}
\allowdisplaybreaks

\theoremstyle{definition}

\newtheorem{theorem}{Theorem}[section]
\newtheorem{corollary}[theorem]{Corollary}
\newtheorem{lemma}[theorem]{Lemma}
\newtheorem{definition}[theorem]{Definition}
\newtheorem{proposition}[theorem]{Proposition}
\newtheorem{remark}[theorem]{Remark}
\newtheorem{notation}[theorem]{Notation}

\newtheorem{example}[theorem]{Example}

\def\K{\mathbb{K}}

\def\F{\mathbb{F}}
\def\Fq{\mathbb{F}_q}

\def\C{\mathcal{C}}
\def\A{\mathcal{A}}
\def\B{\mathcal{B}}
\def\mD{\mathcal{D}}

\def\W{\mathcal{W}}
\def\Z{\mathbb{Z}}

\def\<{\left<}
\def\>{\right>}
\def\trk{\textup{trk}}

\def\maxrk{\textup{maxrk}}
\def\rk{\textup{rk}}
\newcommand{\qbin}[2]{\begin{bmatrix}{#1}\\ {#2}\end{bmatrix}_q}
\newcommand{\bbin}[3]{\begin{bmatrix}{#1}\\ {#2}\end{bmatrix}_{q^{#3}}}

\def\oB{\overline{B}}

\def\omA{\overline{\mathcal{A}}}

\def\maxrk{\textup{maxrk}}

\newcommand{\dimq}[1]{\dim_{\Fq}\left(#1\right)}
\def\oW{\overline{W}}
\def\omW{\overline{\W}}
\def\mP{\mathscr{P}}
\def\ps{\textup{ps}}
\def\cl{\textup{cl}}

\def\D{\textup{D}}
\def\R{\textup{R}}

\def\Q{\mathbb{Q}}
\def\mL{\mathscr{L}}
\def\mM{\mathscr{M}}

\def\PP{\mathcal{P}}
\def\a{\mathfrak{a}}
\def\b{\mathfrak{b}}
\def\mfN{\mathfrak{N}}
\def\n{\mathfrak{n}}
\def\bdX{\boldsymbol{X}}
\def\bdY{\boldsymbol{Y}}
\def\bdT{\boldsymbol{T}}
\def\o{\mathfrak{o}}

\def\c{\mathfrak{c}}
\def\e{\mathfrak{e}}

\def\mfZ{\mathfrak{Z}}
\def\[{\llbracket}
\def\]{\rrbracket}
\def\P{\textup{P}}
\def\mQ{\mathcal{Q}}
\def\dd{\textup{dd}}

\makeatletter
\def\moverlay{\mathpalette\mov@rlay}
\def\mov@rlay#1#2{\leavevmode\vtop{%
   \baselineskip\z@skip \lineskiplimit-\maxdimen
   \ialign{\hfil$\m@th#1##$\hfil\cr#2\crcr}}}
\newcommand{\charfusion}[3][\mathord]{
    #1{\ifx#1\mathop\vphantom{#2}\fi
        \mathpalette\mov@rlay{#2\cr#3}
      }
    \ifx#1\mathop\expandafter\displaylimits\fi}
\makeatother

\newcommand{\bigcupdot}{\charfusion[\mathop]{\bigcup}{\cdot}}

\def\cov{\mathrel{<\kern-.5em\raise.0ex\hbox{$\cdot$}}}

\definecolor{myyellow}{RGB}{218,170,0}
\definecolor{myblue}{RGB}{0,75,135}
\definecolor{mygreen}{RGB}{61,174,43}

\DeclareCaptionLabelSeparator{none}{ }
\title{Zeta Functions for Tensor Codes}

\usepackage{authblk}
\author{Giuseppe Cotardo\thanks{G. C. was supported by the Irish Research Council, grant n. GOIPG/2018/2534.}}
\affil{Department of Mathematics, Virginia Tech, U.S.A}

\usepackage{setspace}
\makeatletter
\newcommand{\subjclass}[2][1991]{%
  \let\@oldtitle\@title%
  \gdef\@title{\@oldtitle\footnotetext{#1 \emph{Mathematics subject classification.} #2}}%
}
\newcommand{\keywords}[1]{%
  \let\@@oldtitle\@title%
  \gdef\@title{\@@oldtitle\footnotetext{\emph{Key words and phrases.} #1.}}%
}
\makeatother

\date{}

\subjclass[2020]{94B05, 15A72}

\keywords{tensor codes; anticodes; tensor weight;  weight enumerator; zeta function; bound}

\begin{document}
\maketitle

\begin{abstract}
    In this work we introduce a new class of optimal tensor codes related to the Ravagnani-type anticodes, namely the $j$-tensor maximum rank distance codes. We show that it extends the family of $j$-maximum rank distance codes and contains the $j$-tensor binomial moment determined codes (with respect to the Ravagnani-type anticodes)  as a proper subclass.  We define and study the generalized zeta function for tensor codes. We establish connections between this object and the weight enumerator of a code with respect to the Ravagnani-type anticodes. We introduce a new refinement of the invariants of tensor codes exploiting the structure of product lattices of some classes of anticodes and we derive the corresponding MacWilliams identities. In this framework, we also define a multivariate version of the tensor weight enumerator and we establish relations with the corresponding zeta function. As an application we derive connections on the generalized tensor weights related to the Delsarte and Ravagnani-type anticodes. 
\end{abstract}

\section{Introduction}

Tensor codes are linear $\Fq$-linear subspaces of the tensor space $\Fq^{n_1}\otimes\cdots\otimes\Fq^{n_r}$ endowed with the tensor rank as a distance function. They were introduced by Roth~\cite{roth1991maximum,roth1996tensor} for the case $n_1=\cdots=n_r$ as a natural generalization of matrix rank-metric codes and used to show that the behaviour of matrix spaces is quite unique compared to the more general case of space of higher order tensors. The definition was then extended in~\cite{byrne2021tensor} wherein four different types of anticodes were classified and exploited to study invariants of tensor codes, such as the tensor weight distribution, the tensor binomial moments and the generalized tensor weights. Although the theory of matrix codes has been the subject of intensive research over the last years, the study of tensor codes is still rather unexplored. In~\cite{roth1991maximum}, Roth provided a generalization of the `usual' Singleton bound which turned out to not be sharp for $r>2$. Even though an improvement of this bound was derived for $3$-tensors, only asymptotic bounds are known for higher order tensors.  Moreover, few classes of tensor codes are known (in fact only those proposed by Roth in~\cite{roth1991maximum,roth1996tensor}). Investigating invariants of these codes represents a crucial step for a better understanding of structural properties and constructing new families of such codes. 

Recently, matrix rank-metric codes were investigated in the context of $3$-tensors~\cite{bartoli2022non,byrne2021bilinear,byrne2019tensor} and the \textit{tensor ranks} of some classes were determined. The tensor rank of a code is defined as the smallest dimension of a space spanned by rank-$1$ tensors containing the code. Another object that captures properties of a code is its \textit{zeta function}. This object was introduced in~\cite{duursma1999weight,duursma2001weight} for Hamming-metric codes and in~\cite{blanco2018rank, byrne2020rank} for matrix rank-metric codes as the generator function of the \textit{generalized normalized binomial moments}. A central role in this theory is played by codes that are extremal with respect to the generalized Singleton-bounds, namely the \textit{$j$-maximum distance separable} ($j$-MDS in short) codes for the Hamming metric case and the \textit{$j$-maximum rank distance} ($j$-MRD in short) codes for the rank-metric case (see~\cite{wei1991generalized} and~\cite{byrne2020rank} respectively for further details). It is well-known that the set of the weight enumerators of such extremal codes provides a $\Q$-basis for the space of homogeneous polynomials in $\Q[X,Y]$ of fixed degree. This property was used in the study of the zeta functions for codes wherein relations between the weight enumerator of a code and the Singleton-optimal weight enumerators were established. In~\cite{byrne2020rank} (for the Hamming and matrix rank-metric case) and in~\cite{byrne2021tensor} (for the tensor case), the family of \textit{$j$-binomial moment determined} ($j$-BMD in short) codes has been investigated. These extremal codes have the property that all or some of their invariants are determined by the code parameters. In particular, in~\cite{byrne2021tensor} the authors showed that the class of $1$-BMD matrix codes is partitioned by the well known families of \textit{MRD} and \textit{dually quasi-MRD} codes (see~\cite{delsarte1978bilinear} and~\cite{de2018weight} for further details). 

In this paper, we set the theory of zeta functions in the most general context of tensor codes with a perspective based on the families of anticodes claasified in~\cite{byrne2021tensor}. In the first part we introduce the notion of zeta function of tensor codes. We then specialize it to the class of Ravagnani-type anticodes and we establish connections between this object and the generalized weight enumerators of tensor codes. As a consequence, we extend the theory developed in~\cite{byrne2020rank} for codes of matrices in the rank metric. The second part will be devoted to a refinement of the tensor zeta functions, binomial moments and weight distributions for closure-type anticodes using the underlying structure of their direct product lattice. We derive analogue relations on these invariants as in the first part of the work.

\textbf{Outline.} The paper is organized as follows. In Section~\ref{sec:prelim} we recall some definitions and results. In Section~\ref{sec:TMRD} we classify and study a new class of tensor codes that are extremal with respect to the Ravagnani-type anticodes, namely the $j$-tensor maximum rank distance codes. We establish relations between this family and that of $j$-tensor BMD codes. We prove that the latter forms a proper subclass of the former. In Section~\ref{sec:zeta} we introduce the tensor zeta function for the different sets of anticodes classified in~\cite{byrne2021tensor}. In particular, we show that this object strictly depends on the underlying set of anticodes. In Section~\ref{sec:zetaR}, we restrict to invariants related to the Ravagnani-type anticodes which allows us to generalize the theory developed in~\cite{byrne2020rank} for matrix codes. In this setting, we relate the tensor zeta function to the tensor weight enumerator and we compute the coefficients of the latter in terms of the weight enumerators of $j$-tensor maximum rank distance codes. The theory in this section can be seen as a generalization of the one in~\cite{byrne2020rank}. Section~\ref{sec:refined} will be devoted to the study of new refined invariants of tensor codes using the fact that the set closure-type anticodes is indeed a \textit{direct product} lattice. More in detail, we define a multivariate version of the tensor weight enumerator and zeta function, and we show connections between these objects. As an application, in Section~\ref{sec:bounds}, we derive connections on the tensor weights associated to the Delsarte and Ravagnani-type anticodes.


\section{Preliminaries and Notation}
\label{sec:prelim}

We start recalling some basic definition on posets and lattices (a standard reference is~\cite[Chapter~3]{stanley2011enumerative}). For any positive integer $i$, we denote by $[i]$ the set $\{1,\ldots,i\}$.

\begin{definition}
	A \textbf{partially ordered set} (\textbf{poset} for short) is a pair $(\mP,\leq)$ where $\mP$ is a non-empty set and $\leq$ is a binary operation satisfying the axioms of reflexivity, antisymmetry and transitivity. We sometimes abuse the notation and write $\mP$ for $(\mP,\leq)$.
\end{definition}

For $s,t\in\mP$ we write $s<t$ for $s\leq t$ and $s\neq t$. Two posets $\mP$ and $\mQ$ are \textbf{isomorphic} if there exists an order-preserving map $\phi:\mP\longrightarrow\mQ$ whose inverse is order preserving, that is $s\leq t \textup{ in }\mP\Longleftrightarrow \phi(s)\leq\phi(t)\textup{ in }\mQ$.

\begin{definition}
	Let $\mP_1,\ldots,\mP_r$ be posets. The \textbf{direct product} of $\mP_1,\ldots,\mP_r$ is the poset $\mP_1\times\ldots\times\mP_r$ on the set $\{(s_1,\ldots,s_r):s_i\in\mP_i$ and $i\in[r]\}$ such that $(s_1,\ldots,s_r)\leq(t_1,\ldots,t_r)$ in $\mP_1\times\ldots\times\mP_r$ if $s_i\leq t_i$ in $\mP_i$ for all $i\in[r]$.
\end{definition}

\begin{definition}
	The \textbf{join} of $s,t\in\mP$, should it exist, is the element $u\in\mP$ satisfying $s\leq u$, $t\leq u$ and $u\leq v$ for all $v\in\mP$ such that $s\leq v$ and $t\leq v$. The \textbf{meet} of $s,t\in\mP$, should it exist, is defined dually as the element $u\in\mP$ satisfying $u\leq s$, $u\leq t$ and $v\leq u$ for all $v\in\mP$ such that $v\leq s$ and $v\leq t$. 
\end{definition}

\begin{notation}
	It easy to check that the join (resp., the meet) of $s,t\in\mP$, should it exist, is unique and we denote it by $s\vee t$ (resp., $s\wedge t$).	
\end{notation}

\begin{definition}
	Let $\mP$ be a poset and let $s,t\in\mP$. The \textbf{interval} $[s,t]$ in $\mP$ is the set $\{u\in\mP:s\leq u\leq t\}$. The poset $\mP$ is \textbf{locally finite} if each of its intervals has only finitely many elements.
\end{definition} 

\begin{definition}
	The \textbf{M\"obius function} of a locally finite poset $\mP$ is defined recursively by $\mu_\mP(s,t)=1$ if $s=t$, $\mu_\mP(s,t)=-\sum_{s\leq u< t}\mu_\mP(s,u)$ if $s<t$ and $\mu_\mP(s,t)=0$ otherwise. We sometimes write $\mu(s,t)$ for $\mu_\mP(s,t)$ if the poset is clear from context.
\end{definition}

\begin{proposition}[M\"obius Inversion Formula]
	Let $\K$ be a field, $\mP$ be a locally finite poset and let $f:\mP\longrightarrow\K$ be a function. Define $g:\mP\longrightarrow\K$ by $g(t)=\sum_{s\leq t}f(s)$ for all $t\in\mP$. Then
	\begin{equation*}
		f(t)=\sum_{s\leq t}\mu_\mP(s,t)g(s) \quad \textup{for all $t \in \mP$}.
	\end{equation*}
\end{proposition}

\begin{proposition}[The Product Theorem]
	Let $\mP_1\times\ldots\times\mP_r$ be the direct product of the locally finite posets $\mP_1,\ldots,\mP_r$. If $(s_1,\ldots,s_r)\leq(t_1,\ldots,t_r)$ in $\mP_1\times\ldots\times\mP_r$ then
	\begin{equation*}
		\mu_{\mP_1\times\ldots\times\mP_r}((s_1,\ldots,s_r),(t_1,\ldots,t_r))=\prod_{i=1}^r\mu_{\mP_i}(s_i,t_i).
	\end{equation*}
\end{proposition}

\begin{definition}
	A \textbf{lattice} $(\mL,\leq)$ is a poset where every $s,t\in\mP$ have join and meet. In particular, the join (resp., the meet) of a non-empty subset $S\subseteq\mL$ is well-defined as the join (resp., the meet) of its elements and denoted by $\bigvee S$ (resp., $\bigwedge S$). A subset $\mM$ of $\mL$ is a \textbf{sublattice} of $\mL$ if $\mM$ is closed under the operations of $\vee$ and $\wedge$ in $\mL$.
\end{definition}

Throughout the paper we let $q$ be a power prime. We let $\Fq$ be the field of cardinality $q$ and let $\F$ be the tensor product space $\Fq^{n_1}\otimes\cdots\otimes\Fq^{n_r}$ for some integers $r,n_1,\ldots,n_r$ such that $r,n_1\geq 2$. We assume $n_1=\min\{n_1,\ldots,n_r\}$ and $n_r=\max\{n_1,\ldots,n_r\}$ without loss of generality and we denote by $n$ the product $n_1\cdots n_r$.

\begin{definition}
	An $r$-tensor $X\in\F$ is a \textbf{rank}-1 (or \textbf{simple}) \textbf{tensor} if $X$ can be expressed as $x^{(1)}\otimes\cdots\otimes x^{(r)}$, for some $x^{(i)}\in\Fq^{n_i}$, $i\in[r]$.
\end{definition}

We say that $V\leq\F$ is \textbf{perfect} if $V$ has a basis of simple tensors (c.f.~\cite{atkinson1983ranks}).

\begin{definition}
	The \textbf{rank} of $X\in\F$ is defined to be the least integer $R$ such that $X$ can be expressed as sum of $R$ simple tensors, that is
	\begin{equation*}
		\rk(X):=\left\{R\in\Z:X=\sum_{j=1}^Rx_j^{(1)}\otimes\cdots\otimes x_j^{(r)}\right\}.
	\end{equation*}
\end{definition}

The rank function induces a metric on $\F$. More in detail, the function $\F\longrightarrow\Z:X\longmapsto\rk(X)$ is a distance function (see~\cite{roth1996tensor}). A basis for $\F$ is given by
\begin{equation*}
	\left\{\bigotimes_{i=1}^ru_{j_i}^{(i)}:j_i\in[n_i] \textup{ and } i\in[r]\right\}
\end{equation*}
provided that $\{u_1^{(i)},\ldots,u_{n_i}^{(i)}\}$ is a basis for $\Fq^{n_i}$, for any $i\in[r]$, and therefore we have  $\dimq{\F}=\prod_{i=1}^r\dimq{\Fq^{n_i}}=n$. We recall that an $r$-tensor $X\in\F$ can be represented as the map $X: [n_1]\times\cdots\times[n_r]\longmapsto \Fq$ given by $X=(X_{j_1,\ldots,j_r}:j_i\in [n_i], i\in [r])$. Therefore, the map 
    \begin{equation*}
        \F\longrightarrow\Fq^{n_1\times\cdots\times n_r}:X=\sum_{s=1}^{\rk(X)}\bigotimes_{i=1}^r x_s^{(i)}\longmapsto \left(X_{j_1,\ldots,j_r}=\sum_{s=1}^{\rk(X)}x_{j_i,s}^{(i)}\right)_{j_i\in [n_i],\; i\in [r]}
    \end{equation*}
   where, $x_s^{(i)}=(x_{j_i,s}^{(i)}:j_i\in [n_i])$ for any $i\in[r]$,  is an $\Fq$-isomorphism. As a consequence, we can identify $\F$ with $\Fq^{n_1\times\cdots\times n_r}$ and represent an $r$-tensor as an $r$-dimensional array.

Tensor codes were introduced in~\cite{roth1991maximum,roth1996tensor} for the case $n_1=n_2=\cdots=n_r$ as a generalization of codes in the rank-metric. We recall the more general definition given in~\cite{byrne2021tensor}.

\begin{definition}
	A \textbf{tensor code} is a $k$-dimensional subspace $\C$ of $\F$. The \textbf{maximum rank} $\maxrk(\C)$ of $\C$ is $\maxrk(\C):=\max\{\rk(C):C\in \C\}$.  The \textbf{tensor rank} $\trk(\C)$ of $\C$ is the minimum dimension of a perfect space containing $\C$. The \textbf{minimum} (\textbf{tensor rank}) \textbf{distance} of a non-zero code $\C$ is $d(\C):=\min\{\rk(C):C \in\C, C\neq 0\}$.
\end{definition}

We refer to $n_1\times\cdots\times n_r,k,d(\C)$ as the \textbf{code parameters} of $\C$ and we say that $\C$ is an $\Fq$-$[n_1\times\cdots\times n_r,k,d(\C)]$ code. Moreover, we simply write $\Fq$-$[n_1\times\cdots\times n_r,k]$ if the minimum distance is not known.  For any $i\in[r]$, we let $\cdot$ be the dot product on $\Fq^{n_i}$, defined in the usual way as $x\cdot y = \sum_{j=1}^{n_i} x_jy_j$ for $x,y \in \Fq^{n_i}$.  We furthermore denote by $*$ the non-degenerate bilinear form defined by
	\begin{equation*}
    	X*Y=\sum_{j_1=1}^{n_1}\cdots\sum_{j_r=1}^{n_r}X_{j_1,\cdots,j_r}Y_{j_1,\cdots,j_r}.
    \end{equation*}

\begin{remark}
	One can check that if $X$ and $Y$ are have rank $1$, say $X=x^(1)\times x^{(r)}$ and $Y=y^(1)\times y^{(r)}$ $*$ then we have
	\begin{equation*}
    	X*Y=\prod_{i=1}^r\left(x^{(i)}\cdot y^{(i)}\right).
    \end{equation*} 
\end{remark}

\begin{definition}
	The \textbf{dual} of $\C$ is defined to be the $\Fq$-subspace of $\F$ which is orthogonal to $\C$ with respect to $*$, that is  $\C^\perp:=\{X\in\F:X*C=0 \textup{ for all } C\in\C\}$.
\end{definition}
    
We summarize some basic facts on the dual tensor code.
    
\begin{lemma}
\label{lem:propdual}
Let $\C,\mD\leq\F$ be codes. The following hold.
	\begin{enumerate}[label=(\arabic*)]
	\setlength\itemsep{0.5em}
	\item\label{item1:propdual} $\left(\C^\perp\right)^\perp=\C$.
	\item\label{item2:propdual} $\dimq{\C^\perp}=n-\dimq{\C}$.
	\item\label{item3:propdual} $(\C\cap\mD)^\perp=\C^\perp+\mD^\perp$.
	\end{enumerate}
\end{lemma}
    
\begin{notation}
	For ease of notation, in the examples throughout the paper we represent $3$-tensors in $\Fq^{n_1}\otimes\Fq^{n_2}\otimes\Fq^{n_3}$ as vectors in $\Fq^{n_1 n_2n_3}$ according to the isomorphism
	\begin{equation*}
	\begin{array}{cccccc}
		&\Fq^{n_1}\otimes\Fq^{n_2}\otimes\Fq^{n_3}&\longrightarrow&\left(\Fq^{n_1\times n_2}\right)^{n_3}&\longrightarrow &\Fq^{n_1n_2n_3}\\[5pt]
		& X &\longmapsto &\left(X_1\mid \cdots \mid X_{n_3}\right) & \longmapsto & \left(\overline{X_1}\mid \cdots \mid \overline{X_{n_3}}\right).
	\end{array}
	\end{equation*} 
	where, for a matrix $M\in\Fq^{n_1\times n_2}$, $\overline{M}$ is the vector of length $n_1n_2$ with entries in $\Fq$ obtained by concatenating the rows of $M$. For example, the $3$-tensor 
	\begin{equation*}
		(1,1)\otimes(1,0,0)\otimes(1,1,1,0)+(1,1)\otimes(0,0,1)\otimes(0,1,1,1)\in\F_2^{2}\otimes\F_2^3\otimes\F_2^4
	\end{equation*}
	can be represented as the vector
	\begin{equation*}
		\arraycolsep=4pt\left(\begin{array}{*{3}{*{3}c;{3pt/3pt}*{3}c|}*{3}c;{3pt/3pt}*{3}c}
			1&0&0&1&0&0&1&0&1&1&0&1&1&0&1&1&0&1&0&0&1&0&0&1
		\end{array}
		\right)\in\F_2^{24}.
	\end{equation*}
\end{notation}

\begin{definition}
	For a code $\C\in\Fq^{n_1}\otimes\Fq^{n_2}\otimes\Fq^{n_3}$ of dimension $k$, we say that $G\in\Fq^{k\times n_1n_2n_3}$ is a \textbf{generator matrix} (resp., a \textbf{parity-check matrix}) for $\C$ if the rows of $G$ are the vector representation of a basis of $\C$ (resp., $\C^\perp$). 
\end{definition}

In the remainder of the paper, unless explicitly stated, $\C\leq\F$ will denote a tensor code of dimension $k$ over $\Fq$ and minimum distance $d$. In~\cite{byrne2021tensor}, the general theory of anticodes for tensor codes was developed and invariants related to the latter were studied. We now recall the main definitions and results. 

\begin{definition}[{\cite[Definition~4.1]{byrne2021tensor}}]
	We say that $A\leq\F$ is a (\textbf{tensor}) \textbf{anticode} if $A$ is perfect.  If $\A$ is a collection of anticodes of $\F$ we write $\A_a$ to denote the set of all anticodes of $\A$ of dimension $a$, where $a\in\{0,\ldots,n\}$. If $\{A^\perp:A\in\A\}$ is a collection of anticodes, then we define the collection of \textbf{dual} (\textbf{tensor}) \textbf{anticodes} of $\A$ to be $\omA:=\{A^\perp:A\in\A\}$. 
\end{definition}

The following definition provide a characterization of the different families of anticodes defined in~\cite{byrne2021tensor}. In particular, it summarizes~\cite[Definition~4.2,~Propositions~4.7,~Theorems~4.16~and~4.21]{byrne2021tensor}.
\begin{definition}
\label{def:anticodes}
	We define the following.
	\begin{enumerate}[label=(\arabic*)]
	\setlength\itemsep{3pt}
		\item The family of \textbf{perfect spaces} is\\
			$\displaystyle \A^\ps:=\{A\in\F:A \textup{ is perfect}\}$.
		\item The family of \textbf{closure-type anticodes} is\\
			$\displaystyle \A^\cl:=\left\{\bigotimes_{i=1}^rA^{(i)}:A^{(i)}\leq\Fq^{n_i} \textup{ for any } i\in[r]\right\}$.
		\item The family of \textbf{Delsarte-type anticodes} is\\
			$\displaystyle\A^\D=\bigcup_{p\in P}\left\{\left(\bigotimes_{j=1}^{i-1}\Fq^{n_j}\right)\otimes A^{(i)}\otimes\left(\bigotimes_{j=i+1}^r\Fq^{n_j}\right):A^{(i)}\leq\Fq^{n_i}, i \in [r]\setminus\{p\} \right\}$\\[1pt]
			where $P:=\{i\in[r]:n_i=n_r\}$.
		\item The family of \textbf{Ravagnani-type anticodes} is\\
			$\displaystyle \A^\R=\left\{\left(\bigotimes_{j=1}^{i-1}\Fq^{n_j}\right)\otimes A^{(i)}\otimes\left(\bigotimes_{j=i+1}^r\Fq^{n_j}\right):A^{(i)}\leq\Fq^{n_i}, i \in S \right\}$\\[1pt]
			where $S:=\{i\in[r]:n_i=n_1\}$.
	\end{enumerate}
\end{definition}

It was noted in~\cite{byrne2021tensor} that for $r=2$ the families of Delsarte-type and Ravagnani-type anticodes coincide and reduce to the class of optimal anticodes defined in~\cite{ravagnani2016generalized}. It was also observed in ~\cite[Example~4.5]{byrne2021tensor} that the dual of a perfect space is not perfect in general and therefore $\overline{\A^\ps}\not\subseteq\A^\ps$. The next result summarizes~\cite[Corollaries~4.13,~4.17,~4.22]{byrne2021tensor}.

\begin{proposition}
\label{prop:charoverlineA}
	We have $\overline{\A^\cl},\overline{\A^\D},\overline{\A^\R}\subseteq\A^\ps$. In particular, $\overline{\A^\D}=\A^\D$, $\overline{\A^\R}=\A^\R$ and
	\begin{equation*}
    	\overline{\A^\cl}=\left\{\sum_{i=1}^r\left(\bigotimes_{j=1}^{i-1}\Fq^{n_j}\right)\otimes A^{(i)}\otimes\left(\bigotimes_{j=i+1}^r\Fq^{n_j}\right):A^{(i)}\leq\Fq^{n_i}\textup{ for all } i \in [r]\right\}.
	\end{equation*}
\end{proposition}

\begin{remark}
	We recall that the families of anticodes $\A^\ps,\A^\cl,\overline{\A^\cl},\A^\D,\overline{\A^\D},\A^R$ and $\overline{\A^\R}$ are indeed lattices, as proved in~\cite{byrne2021tensor}.
\end{remark}

The following invariants were introduced in~\cite{byrne2021tensor}. They extend the \textit{Delsarte generalized weights}~\cite{ravagnani2016generalized} and the \textit{generalized tensor ranks} ~\cite{byrne2019tensor} which can be recovered for $r=2$, and $\A=\A^D$ and $\A=\A^\ps$ respectively.

\begin{definition}[{\cite[Definition~5.1]{byrne2021tensor}}]
\label{def:tj}
	Let $\A$ be a collection of anticodes. For each $j\in[k]$, the $j$\textbf{-th tensor weight} with respect to $\A$ is defined to be:
		\begin{equation*}
			t_j(\C):=\min\left\{\dimq{A}:A \in\A \textup{ and } \dimq{\C \cap A}\geq j\right\}.
		\end{equation*}
	Furthermore, if $\omA \subseteq \A^\ps$ then we define the $j$\textbf{-th dual tensor weight} to be
		\begin{equation*}
			s_j(\C):=\min\left\{\dimq{A}:A \in\omA\textup{ and } \dimq{\C \cap A}\geq j\right\}.
		\end{equation*}
\end{definition}

For the remainder, we let $j$ be a fixed integer in $[k]$ and we write $t_j$, $s_j$, $t_j^\perp$ and $s_j^\perp$ instead of $t_j(\C)$, $s_j(\C)$, $t_j\left(\C^\perp\right)$ and $s_j\left(\C^\perp\right)$ respectively.

\begin{proposition}[{\cite[Proposition~5.2]{byrne2021tensor}}]
\label{prop:proptj}
	The following hold.
	\begin{enumerate}[label=(\arabic*)]
	\setlength\itemsep{3pt}
		\item $t_j\leq t_{j+1}$ and, if $\omA\subseteq\A^\ps$, $s_j\leq s_{j+1}$  for any $j\in[k]$.
		\item $t_j^\ps\leq t_j^\cl\leq t_j^\D\leq t_j^\R$ for any $j\in[k]$ .
		\item $t_j^\D=s_j^\D$ and $t_j^\R=s_j^\R$ for any $j\in[k]$.
	\end{enumerate}
\end{proposition}

In~\cite[Section~6]{byrne2021tensor}, new invariants related to the different collections of anticodes were defined, namely the \textit{tensor binomial moments} and \textit{weight distributions} and it was shown that they encode the same information. We briefly recall their definitions and some main results. We use the convention that $\sum_{x\in\emptyset}f(x)=0$ for any function $f:\Q\longrightarrow\Q$.

\begin{definition}[{\cite[Definition~6.1]{byrne2021tensor}}]
    For any $a\in\{0,\ldots,n\}$ the $(a,j)$\textbf{-th tensor binomial moment} of $\C$ is defined to be
    \begin{equation*}
    	B_a^{(j)}(\C):=\sum_{A\in\A_a}B_A^{(j)}(\C),\qquad\textup{ where }\qquad B_A^{(j)}(\C):=\qbin{\dimq{\C\cap A}}{j}
	\end{equation*}
	for any $A\in \A$. If moreover $\omA\subseteq\A^\ps$ we define the $(a,j)$\textbf{-th dual tensor binomial moment} of $\C$ to be
	\begin{equation*}
    	\oB_a^{(j)}(\C):=\sum_{A\in\omA_a}\oB_{A}^{(j)}(\C),\qquad\textup{ where }\qquad \oB_{A}^{(j)}(\C):=\qbin{\dimq{\C\cap A}}{j}.
	\end{equation*}
 \end{definition}
  
\begin{definition}[{\cite[Definition~6.2]{byrne2021tensor}}]
	 The $j$\textbf{-th weight distribution} is defined to be the vector of length $n+1$ whose $(a+1)$-th component is given by
	\begin{equation*}
    	W_a^{(j)}(\C):=\sum_{A\in\A_a}W_A^{(j)}(\C),
	\end{equation*}
	where $W_A^{(j)}:=|\{\mD\leq(\C\cap A):\dimq{\mD}=j \textup{ and } \mD\leq B\leq A, B\in\A \Longrightarrow B=A\}|$, for any $A\in\A_a$. If  $\omA\subseteq\A^\ps$ we let the $j$\textbf{-th dual weight distribution} to be the vector of length $n+1$ whose $(a+1)$-th component is given by
	\begin{equation*}
    	\oW_a^{(j)}(\C):=\sum_{A\in\omA_a}\oW_A^{(j)}(\C),
	\end{equation*}
	where $\oW_A^{(j)}:=|\{\mD\leq(\C\cap A):\dimq{\mD}=j \textup{ and } \mD\leq B\leq A, B\in\omA \Longrightarrow B=A\}\}|$, for any $A\in\omA_a$.
\end{definition}
  	
\begin{definition}[{\cite[Definition~6.3]{byrne2021tensor}}]
	Let $X$ and $Y$ be indeterminates. We define the $j$\textbf{-th tensor weight enumerator} by
	\begin{equation*}
    	\W_\C^{(j)}(X,Y):=\sum_{a\leq n}W_a^{(j)}(\C)X^{n-a}Y^a\in \Q[X,Y].
	\end{equation*}
	If  $\omA\subseteq\A^\ps$ we let the $j$\textbf{-th dual tensor weight enumerator} be defined by
	\begin{equation*}
    	\omW_\C^{(j)}(X,Y):=\sum_{a\leq n}\oW_a^{(j)}(\C)X^{n-a}Y^a\in \Q[X,Y].
	\end{equation*}
\end{definition}

\begin{proposition}[{\cite[Theorem~6.4]{byrne2021tensor}}]
\label{prop:oldBaWb}
	Let $\A$ be a family of anticodes whose members form a lattice and let $\mu$ be its M\"obius function. The following hold for all $a,b\in\{0,\ldots,n\}$.
	\begin{enumerate}[label=(\arabic*)]
  		\item $\displaystyle B_a^{(j)}(\C)=\sum_{b=0}^aW_b^{(j)}(\C)\;\left|\left\{(A',A)\in\A_b\times\A_a:A'\leq A\right\}\right|$.
    	\item $\displaystyle W_{b}^{(j)}(\C)=\sum_{a=0}^b\mu(a,b)B_a^{(j)}(\C)\;\left|\left\{(A',A)\in\A_b\times\A_a:A\leq A'\right\}\right|$.
	\end{enumerate}
\end{proposition} 

\begin{proposition}[{\cite[Theorem~6.6]{byrne2021tensor}}]
\label{prop:Badetermied}
    	Suppose that $\omA\subseteq\A^\ps$. For any $a\in\{0,\ldots,n\}$, we have
	\begin{equation*}
		B_a^{(j)}(\C)=
		\begin{cases}
    		0 & \textup{ if } a < t_j,\\[1ex]
			\qbin{k+a-n}{j}\left|\A_a\right| & \textup{ if } a > n-s_1^\perp.
	\end{cases}
	\end{equation*}  
\end{proposition}

A family of extremal codes with the property that the tensor binomial moments and weight distribution are partially determined by their code parameters was introduced in~\cite{byrne2020rank} and is defined as follows.

\begin{definition}[{\cite[Definition~6.9]{byrne2021tensor}}]
    Suppose that  $\omA\subseteq\A^\ps$. We say that the code $\C$ is  $j$-\textbf{tensor binomial moment determined} ($j$\textbf{-TBMD} in short) with respect to (the anticodes in)  $\A$ if $n-s_1^\perp-t_j<0$. Moreover, we say that $\C$ is \textbf{minimally} $j$\textbf{-TBMD} if $j$ is the minimum of the set $\{p\in[k]:\C \textup{ is } p\textup{ -TBMD}\}$.
\end{definition}

Note that, this notion of extremality generalizes the one of being $j$-BMD (see~\cite{byrne2020rank}), which can be recovered for $r=2$ and $\A=\A^\D=\A^\R$. Finally, we recall some well-known  properties of $q$-binomial coefficients (the reader is referred to~\cite{andrews1998theory} for more details).
	
	\begin{definition}
		Let $a,b$ be integers. The $q$-binomial coefficient of $a$ and $b$ is defined to be 
		\begin{equation*}
			\qbin{a}{b}:=
			\begin{cases}
				0 & \textup{ if } b<0 \textup{ or } 0\leq a \leq b,\\[1ex]
				1 & \textup{ if } b=0,\\[1ex]
				\displaystyle\prod_{i=0}^{b-1}\frac{q^{a-i}-1}{q^{i+1}-1} & \textup{ if } b>0 \textup{ and } a\geq b.
			\end{cases}
		\end{equation*}
	\end{definition}
	
	\begin{lemma}
		\label{lem:bin}
		Let $a,b,c$ be integers and $x,y$ be rational numbers. The following hold.
		\vspace{5pt}
		\begin{enumerate}[label=(\arabic*)]
		\setlength\itemsep{5pt}
			\item $\displaystyle \qbin{a}{b}\qbin{b}{c}=\qbin{a}{c}\qbin{a-c}{a-b}$.
			\item $\displaystyle \qbin{a+b}{c}=\sum_{j=0}^cq^{j(b-c+j)}\qbin{a}{j}\qbin{b}{c-j}=\sum_{j=0}^cq^{(c-j)(a-j)}\qbin{a}{j}\qbin{b}{c-j}$.
			\item $\displaystyle\sum_{j=0}^c\qbin{c}{j}(-1)^jq^{\binom{j}{2}}x^{c-j}y^j=\begin{cases}
				0 & \textup{ if } c=0,\\
				\displaystyle\prod_{j=0}^{c-1}(x-q^jy) & \textup{ if } c\geq 1.
			\end{cases}$
		\end{enumerate} 
	\end{lemma}


\section{$j$-TMRD Codes for Ravagnani-type Anticodes}
\label{sec:TMRD}
	In this section, we introduce a new class of codes that show properties of extremality  with respect to the Ravagnani-type anticodes and we relate them to the $j$-TBMD codes. We prove that the latter are a proper subclass of the former and we show that they extend the class of $1$-MRD matrix rank-metric codes. We recall some results proved in~\cite{byrne2021tensor} on the generalized tensor weight related to $\A^\R$.

	\begin{proposition}[{\cite[Proposition~5.10]{byrne2021tensor}}]
\label{prop:proptR}
	The following hold. 
	\begin{enumerate}[label=(\arabic*)]
		\item\label{item1:proptR}  $t_j^\R +\frac{n}{n_1}\leq t_{j+\frac{n}{n_1}}^\R$. 
 		\item\label{item2:proptR} $t_j^\R\leq n-\frac{n}{n_1}\left\lfloor\frac{n_1}{n}(k-j)\right\rfloor $. 
		\item\label{item3:proptR} $t_j^\R\geq j$.
	\end{enumerate}
\end{proposition}

As observed in~\cite{byrne2021tensor}, the Wei-type duality described in~\cite[Section~6]{ravagnani2016generalized} can be extended for the tensor weights associated to the Ravagnani-type anticodes.    	
   	\begin{theorem}[{\cite[Theorem~5.12]{byrne2021tensor}}]
   	\label{thm:weiduality}
   		Define the sets
        \begin{align*}
            S_p(\C^\perp)&:=\left\{\frac{n_1}{n}\,t_{p+i\frac{n}{n_1}}^\R(\C^\perp):i\in\Z \textup{ and } 1\leq p+i\frac{n}{n_1}\leq k\right\},\\ \overline{S}_p(\C)&:=\left\{n_1+1-\frac{n_1}{n}\,t_{p+i\frac{n}{n_1}}^\R(\C):i\in\Z \textup{ and }1\leq p+i\frac{n}{n_1}\leq k\right\}.
        \end{align*}
        We have $S_p(\C^\perp)=[n_1]\setminus\overline{S}_{p+k}(\C)$ for any $1\leq p\leq \frac{n}{n_1}$. In particular, the tensor weights (of the Ravagnani-type) of $\C$ fully determine those of $\C^\perp$.
   	\end{theorem}
   	
In the remainder, we set $t_1(\{0\}):=n+\frac{n}{n_1}$. This choice is in line with~\cite[Definition~11.3.1]{huffman2021concise} as we will observe after Lemma~\ref{lem:CCdual1TMRD}. We introduce the main object of this section, that is a new class of codes meeting the bound in Proposition~\ref{prop:proptR}\eqref{item2:proptR} with equality. This bound is a possible generalization of the Singleton-type bound for matrix codes.
   	
   	\begin{definition}
   		The code $\C$ is $j$-\textbf{tensor maximum rank distance} ($j$-\textbf{TMRD} in short) with respect (the anticodes in) $\A^\R$ if
   		\begin{equation*}
   			t_j^\R=n-\frac{n}{n_1}\left\lfloor\frac{n_1}{n}(k-j)\right\rfloor.
   		\end{equation*}
   	\end{definition}

\begin{remark}
	It is immediate to see that the property of being $j$-TMRD with respect to $\A^\R$ extends the one of being $j$-MRD in the matrix case (see~\cite[Definition~4.6]{byrne2020rank}). In particular, a code $\C\leq\Fq^{n_1}\otimes\Fq^{n_2}$ is $j$-TMRD with respect to $\A^\R$ if and only if $\C$ is $j$-MRD.
\end{remark}

\begin{remark}
\label{rem:MRD}
	A code  $\C\leq\Fq^{n_1}\otimes\Fq^{n_2}$ that satisfies
	\begin{equation*}
		d=n_1-\left\lfloor\frac{k-1}{n_2}\right\rfloor
	\end{equation*}
	is said to be $1$-MRD (see~\cite[Definition~4.6]{byrne2020rank}). More in detail, it was observed in~\cite[Remark~4.17]{byrne2020rank} that $\C$ is \textbf{MRD} (\textbf{maximum rank distance}) if $n_2\mid k$ and $\C$ is \textbf{QMRD} (\textbf{quasi maximum rank distance}) if $n_2\nmid k$. These codes exist for any value of $n_1,n_2$ and $k$ (see~\cite{de2018weight,delsarte1978bilinear} for further details). Clearly, we have that if $r=2$ then $\C$ is $1$-TMRD with respect $\A^\R$ if and only if $\C$ is MRD or QMRD, since $t_1^\R=d\,n_2$ in this case.
\end{remark}

A code of matrices $\C$ is $j$-MRD then it is $(j+n_2)$-MRD by~\cite[Lemma~4.8]{byrne2020rank}. We extend this result for $r>2$.
    
\begin{lemma}
\label{lem:j+mTBMD}
	If $\C$ is $j$-TMRD  with respect $\A^\R$ then $\C$ is $(j+\frac{n}{n_1})$-TMRD with respect $\A^\R$.
\end{lemma}
\begin{proof}
	Suppose that $\C$ is $1$-TMRD  with respect $\A^\R$. By Proposition~\ref{prop:proptR}, we have
	\begin{equation*}
		n-\frac{n}{n_1}\left\lfloor\frac{n_1}{n}(k-j)\right\rfloor+\frac{n}{n_1}=t_j^\R+\frac{n}{n_1}\leq t_{j+\frac{n}{n_1}}^\R\leq n-\frac{n}{n_1}\left\lfloor\frac{n_1}{n}\left(k-j-\frac{n}{n_1}\right)\right\rfloor.
	\end{equation*}
	It is immediate to see that the first and the last terms are equal. The statement follows. 
\end{proof}

The next example shows that a $j$-TMRD code  with respect to $\A^\R$ is not necessarily $(j+1)$-TMRD with respect to $\A^\R$. 

\begin{example}
\label{ex:1-TMRD}
	Let $\C\leq\F_2^2\otimes\F_2^3\otimes\F_2^4$ be the code of dimension $13$ whose generator matrix is
	\begin{equation*}
		\left(\begin{array}{*{3}{*{3}c;{3pt/3pt}*{3}c|}*{3}c;{3pt/3pt}*{3}c}
			1&0&0&0&0&0&0&0&0&0&0&0&0&0&1&1&0&0&1&0&0&0&1&0\\
			0&1&0&0&0&0&0&0&0&0&0&0&0&0&1&0&0&1&0&1&1&1&0&0\\
			0&0&1&0&0&0&0&0&0&1&0&0&0&0&0&1&0&0&0&1&1&1&1&1\\
			0&0&0&1&0&0&0&0&0&1&0&0&0&0&0&1&0&1&0&1&1&1&1&1\\
			0&0&0&0&1&0&0&0&0&1&0&0&0&0&1&1&0&1&1&0&1&1&0&1\\
			0&0&0&0&0&1&0&0&0&1&0&0&0&0&1&0&1&0&0&1&0&1&1&1\\
			0&0&0&0&0&0&1&0&0&0&0&0&0&0&0&0&0&1&1&1&1&0&0&1\\
			0&0&0&0&0&0&0&1&0&1&0&0&0&0&0&1&0&1&0&0&0&0&1&1\\
			0&0&0&0&0&0&0&0&1&0&0&0&0&0&1&0&0&0&1&1&1&1&0&0\\
			0&0&0&0&0&0&0&0&0&0&1&0&0&0&0&1&0&0&1&0&0&0&1&0\\
			0&0&0&0&0&0&0&0&0&0&0&1&0&0&1&0&1&0&0&0&1&0&1&0\\
			0&0&0&0&0&0&0&0&0&0&0&0&1&0&0&0&0&1&1&0&0&0&0&1\\
			0&0&0&0&0&0&0&0&0&0&0&0&0&1&1&0&0&0&1&1&1&1&1&1
		\end{array}\right)
	\end{equation*}
	One can check the following.
	\begin{itemize}
		\item  $t_1^\R(\C)=t_2^\R(\C)=12$. This value is attained, for example, for the Ravagnani-type anticode $\<(1,0)\>_{\F_2}\otimes\F_2^3\otimes\F_2^4$ whose intersection with $\C$ is the row-space of 
			\begin{equation*}
			\left(\begin{array}{*{3}{*{3}c;{3pt/3pt}*{3}c|}*{3}c;{3pt/3pt}*{3}c}
				0&1&1&0&0&0&0&1&0&0&0&0&0&0&1&0&0&0&0&0&0&0&0&0\\
				0&0&0&0&0&0&1&0&0&0&0&0&1&0&0&0&0&0&0&1&1&0&0&0
			\end{array}\right).
			\end{equation*}
		\item $t_3^\R(\C)=\cdots=t_{13}^\R(\C)=24$ which means that the full-space $\F_2^2\otimes\F_2^3\otimes\F_2^4$ is the only Ravagnani-type anticode that intersect the code in a subspace of dimension at least $3$.
	\end{itemize}
	We have
	\begin{equation*}
		t_1^\R(\C)=12=24-12\left\lfloor\frac{13-1}{12}\right\rfloor\quad \textup{ and }\quad t_2^\R(\C)=12\neq24=24-12\left\lfloor\frac{13-2}{12}\right\rfloor
	\end{equation*}
	which implies that $\C$ is $1$-TMRD with respect to $\A^\R$ but not $2$-TMRD with respect to $\A^\R$. Finally
	\begin{equation*}
		t_{13}^\R(\C)=24=24-12\left\lfloor\frac{13-13}{12}\right\rfloor
	\end{equation*}
	and $\C$ is also $13$-TBMD, in line with Lemma~\ref{lem:j+mTBMD}.
\end{example}

The following generalizes~\cite[Lemma~4.11]{byrne2020rank}. The proof is similar but we include it for completeness.

\begin{proposition}
\label{prop:jTMRD}
	We have that $\C$ is $j$-TMRD with respect $\A^\R$ for all $j\in[k]$  if and only if $\C$ is $1$-TMRD  with respect $\A^\R$ and one of the following conditions are satisfied.
	\begin{enumerate}[label=(\arabic*)]
		\item $\frac{n}{n_1}\mid k$.
		\item $\frac{n}{n_1}\nmid k$ and $\C^\perp$ is $1$-TMRD  with respect $\A^\R$.
	\end{enumerate}
\end{proposition}    
    \begin{proof}
    	Write $k=\alpha\frac{n}{n_1}+\rho$ for some $\alpha\in\Z_{\geq 0}$ and $0\leq \rho\leq \frac{n}{n_1}-1$. By Lemma~\ref{lem:j+mTBMD}, it is sufficient to prove that if $\C$ is $1$-TMRD with respect $\A^\R$ than $\C$ is $j$-TMRD with respect $\A^\R$ for all $1\leq j\leq \frac{n}{n_1}$. Suppose $\C$ to be $1$-TMRD with respect $\A^\R$ and let $1\leq j\leq \frac{n}{n_1}$. If $\rho=0$, i.e. $\frac{n}{n_1}\mid k$, then we have
    	\begin{equation*}
    		n-\frac{n}{n_1}\alpha=t_1^\R\leq t_j^{\R}\leq t_{1+\frac{n}{n_1}}^\R-\frac{n}{n_1}=n-\frac{n}{n_1}\alpha
    	\end{equation*}   
    	by Propositions~\ref{prop:proptj} and~\ref{prop:proptR}. This proves the first part of the statement so we assume $\frac{n}{n_1}\nmid k$ and $\C^\perp$ $1$-TMRD with respect $\A^\R$ for the remainder of the proof. We distinguish two cases according to the value of $\rho$.
    	\begin{enumerate}[label=(\arabic*)]
    		\item If $j\leq \rho\leq \frac{n}{n_1}-1$ then Propositions~\ref{prop:proptj} and~\ref{prop:proptR} imply
    			\begin{equation*}
    				n-\frac{n}{n_1}\alpha=t_1^\R\leq t_j^\R\leq t_\rho^\R=n-\frac{n}{n_1}\alpha.
    			\end{equation*}
    		\item If $1\leq\rho<j\leq\frac{n}{n_1}$ then one can easily check that
    		\begin{equation*}
    			t_1^\R=n-\frac{n}{n_1}(\alpha+1)\qquad\textup{ and }\qquad \left(t_1^\R\right)^\perp=\frac{n}{n_1}(\alpha+1).
    		\end{equation*}
    		Observe that, by Propositions~\ref{prop:proptR}, we have 
    		\begin{equation*}
    			\frac{n}{n_1}(\alpha+1)=\left(t_1^\R\right)^\perp<\left(t_{1+\frac{n}{n_1}}^\R\right)^\perp<\cdots<\left(t_{1+(n_1-\alpha-1)\frac{n}{n_1}}^\R\right)^\perp\leq n
    		\end{equation*}
    		which implies 
    		\begin{equation}
    		\label{eq:tRperp}
    			\left(t_{1+s\frac{n}{n_1}}^\R\right)^\perp=\frac{n}{n_1}(\alpha+1+s) 
    		\end{equation}
    		for all $s\in\{0,\ldots, n_1-\alpha-1\}$. Define the sets 
    		\begin{align*}
    			S_1\left(\C^\perp\right)&:=\left\{\frac{n_1}{n}\left(t_1^\R\right)^\perp,\frac{n_1}{n}\left(t_{1+\frac{n}{n_1}}^\R\right)^\perp,\cdots,\frac{n_1}{n}\left(t_{1+(n_1-\alpha-1)\frac{n}{n_1}}^\R\right)^\perp\right\},\\
    			\overline{S}_{1+k}\left(\C\right)&:=\left\{n_1+1-\frac{n_1}{n}t_{k+1-\frac{n}{n_1}}^\R,n_1+1-\frac{n_1}{n}t_{k+1-2\frac{n}{n_1}}^\R,\cdots,n_1+1-\frac{n_1}{n}t_{k+1-\alpha\frac{n}{n_1}}^\R\right\},
    		\end{align*}
    		and recall that, by Theorem~\ref{thm:weiduality}, we have $S_1\left(\C^\perp\right)=[n_1]\setminus	\overline{S}_{1+k}\left(\C\right)$. As a consequence of~\cite[Proposition~5.10]{byrne2021tensor} and Equation~\eqref{eq:tRperp} we have
    		\begin{equation*}
    			n_1+1-\frac{n_1}{n}t_{k+1-\frac{n}{n_1}}^\R< n_1+1-\frac{n_1}{n}t_{k+1-2\frac{n}{n_1}}^\R< \cdots< n_1+1-\frac{n_1}{n}t_{k+1-\alpha\frac{n}{n_1}}^\R= \alpha
    		\end{equation*}
    		which implies
    		\begin{equation*}
    			t_{k+1-\alpha\frac{n}{n_1}}^\R=t_{\rho+1}^\R=n-\frac{n}{n_1}(\alpha+1).
    		\end{equation*}
    		Finally, by~\cite[Proposition~5.10]{byrne2021tensor}, we get
    		\begin{equation*}
    			n-\frac{n}{n_1}(\alpha+1)=t_{\rho+1}^\R\leq t_j^\R\leq t_{\frac{n}{n_1}}^\R\leq n-\frac{n}{n_1}(\alpha+1).
    		\end{equation*}
    	\end{enumerate}   	
    	This concludes the proof.
   	\end{proof}
   	
  	\begin{remark}
   		In Proposition~\ref{prop:jTMRD}, we introduced two classes of codes, namely 
   		\begin{enumerate}[label=(\arabic*)]
   			\item\label{item1:MRD} $\left\{\C\leq\F:\frac{n}{n_1}\mid k\textup{ and }\C \textup{ is $1$-TMRD with respect to }\A^\R\right\}$,
   			\item\label{item2:dQMRD} $\left\{\C\leq\F:\frac{n}{n_1}\nmid k\textup{ and both }\C \textup{ and }\C^\perp\textup{ are $1$-TMRD with respect to }\A^\R\right\}$.
   		\end{enumerate}
   		The first set extends the family of MRD codes as observed in Remark~\ref{rem:MRD}. On the other hand,~\ref{item2:dQMRD} generalizes the class of dually QMRD codes introduced in~\cite{de2018weight} wherein their existence was proved for any choice of $q,n,m$ and $k$.
   	\end{remark}

   	It is well-known that the dual of an MRD code is MRD as well. The next result shows that this property extends also to $1$-TMRD codes with respect to $\A^\R$ whose dimensions are divisible by $\frac{n}{n_1}$.
   	
   	\begin{lemma}
   	\label{lem:CCdual1TMRD}
   		If $\C$ is $1$-TMRD and $\frac{n}{n_1}\mid k$ then $\C^\perp$ is $1$-TMRD. 
   	\end{lemma}
   	\begin{proof}
   		Write $k=\alpha\frac{n}{n_1}$ and define the sets 
   		\begin{align*}
    		S_1\left(\C^\perp\right)&:=\left\{\frac{n_1}{n}\left(t_1^\R\right)^\perp,\frac{n_1}{n}\left(t_{1+\frac{n}{n_1}}^\R\right)^\perp,\ldots,\frac{n_1}{n}\left(t_{1+(n_1-\alpha-1)\frac{n}{n_1}}^\R\right)^\perp\right\}\\
    		\overline{S}_{1+k}\left(\C\right)&:=\left\{n_1+1-\frac{n_1}{n}t_{k+1-\frac{n}{n_1}}^\R,n_1+1-\frac{n_1}{n}t_{k+1-2\frac{n}{n_1}}^\R,\ldots,n_1+1-\frac{n_1}{n}t_{k+1-\alpha\frac{n}{n_1}}^\R\right\}.
    	\end{align*}
    	Proposition~\ref{prop:jTMRD} implies
    	\begin{equation*}
    		n_1+1-\frac{n_1}{n}t_{k+1-s\frac{n}{n_1}}^\R=s
    	\end{equation*}
    	for all $s\in[\alpha]$, and therefore $\left(t_1^\R\right)^\perp=s+1$ by Proposition~\ref{prop:proptR}. Finally, one can observe that
    	\begin{equation*}
    		\alpha+1=\left(t_1^\R\right)^\perp\leq n-\frac{n}{n_1}\left\lfloor\frac{n_1}{n}(n-k-1)\right\rfloor=\frac{n}{n_1}\left\lceil\frac{n_1}{n}(k+1)\right\rceil=\alpha+1
    	\end{equation*}
    	which implies the statement.
   	\end{proof}
   	
  	\begin{remark}
  		One can easily check that $t_1^\R(\F)=\frac{n}{n_1}$ since $d(\F)=1$ and the smallest Ravagnani-type anticode containing a simple $r$-tensor must have dimension $\frac{n}{n_1}$. Moreover, the code $\{0\}$ is the dual of $\F$ and, using this convention $t_1^\R(\{0\})=n+\frac{n}{n_1}$, we have that they are both $1$-TMRD in line with Lemma~\ref{lem:CCdual1TMRD}.
  	\end{remark}
   	
 The next result generalizes~\cite[Corollary~18]{de2018weight}.
    
    \begin{proposition}
    \label{prop:boundt1+t1perp}
    	The following holds.
    	\begin{enumerate}[label=(\arabic*)]
    		\item If $\frac{n}{n_1}\mid k$ then either $t_1^\R+\left(t_1^\R\right)^\perp\leq n$ or $t_1^\R+\left(t_1^\R\right)^\perp=n+2\frac{n}{n_1}$ and $\C$ is $1$-TMRD with respect to $\A^\R$.
    		\item If $\frac{n}{n_1}\nmid k$ then $t_1^\R+\left(t_1^\R\right)^\perp\leq n+\frac{n}{n_1}$ and the equality holds if and only if both $\C$ and $\C^\perp$ are $1$-TMRD with respect to $\A^\R$.
    	\end{enumerate}
    \end{proposition}
    \begin{proof}
    	Write $k=\alpha\frac{n}{n_1}+\rho$ with $\rho\in\left\{0,\ldots,\frac{n}{n_1}-1\right\}$. Proposition~\ref{prop:proptR} implies
    	\begin{equation}
    	\label{eq:t1R}
    		t_1^\R\leq n-\frac{n}{n_1}\left\lfloor\frac{n_1}{n}(k-1)\right\rfloor\quad\textup{ and }\quad \left(t_1^\R\right)^\perp\leq n-\frac{n}{n_1}\left\lfloor\frac{n_1}{n}(k^\perp-1)\right\rfloor=\frac{n}{n_1}\left\lceil\frac{n_1}{n}(k+1)\right\rceil.
    	\end{equation}
    	By summing these expressions, we obtain 
    	\begin{equation*}
    		t_1^\R+\left(t_1^\R\right)^\perp\leq n-\frac{n}{n_1}\left\lfloor\frac{n_1}{n}(\rho-1)\right\rfloor+\frac{n}{n_1}\left\lceil\frac{n_1}{n}(\rho+1)\right\rceil.
    	\end{equation*}
    	If $\rho\neq 0$, i.e. $\frac{n}{n_1}\nmid k$, we have $0\leq\rho-1<\rho+1\leq\frac{n}{n_1}$ which implies $t_1^\R+\left(t_1^\R\right)^\perp\leq n+\frac{n}{n_1}$. In addition, the equality holds if and only if the inequalities in~\eqref{eq:t1R} are equalities, that is if and only if $\C$ and $\C^\perp$ are $1$-TMRD with respect to $\A^\R$. We assume $\rho=0$, i.e. $\frac{n}{n_1}\mid k$, in the remainder of the proof. We have $t_1^\R+\left(t_1^\R\right)^\perp\leq n+2\frac{n}{n_1}$. Observe that the $t_1^\R+\left(t_1^\R\right)^\perp= n+2\frac{n}{n_1}$ if and only if the inequality in~\eqref{eq:t1R} are equalities, that is if and only if $\C$ is $1$-TBMD. Finally, suppose that the equality does not hold for one of the expressions in~\eqref{eq:t1R} then, by  Lemma~\ref{lem:CCdual1TMRD}, the inequality cannot hold for the other one. Thus, if $\C$ is not $1$-TBMD then 
    	\begin{equation}
    	\label{eq:t1Rno1TMRD}
    		t_1^\R\leq n-\frac{n}{n_1}\left\lfloor\frac{n_1}{n}(k-1)\right\rfloor-\frac{n}{n_1}\quad\textup{ and }\quad \left(t_1^\R\right)^\perp\leq\frac{n}{n_1}\left\lceil\frac{n_1}{n}(k+1)\right\rceil-\frac{n}{n_1}
    	\end{equation}
    		since the dimension of a Ravagnani-type anticode is a multiple of $\frac{n}{n_1}$. The statement now follows by summing the inequalities in~\eqref{eq:t1Rno1TMRD}.
   	\end{proof}
   	
\begin{example}
\label{ex:1-TMRDm|k}
	Let $\C$ be the $\F_2$-$[3\times 3\times 4,12]$ code generated by 
	\begin{equation*}
	\resizebox{\textwidth}{!}{$
		\left(\begin{array}{*{3}{*{3}c;{3pt/3pt}*{3}c;{3pt/3pt}*{3}c|}*{3}c;{3pt/3pt}*{3}c;{3pt/3pt}*{3}c}
	1&0&0&0&0&0&0&0&0&0&0&0&0&0&1&0&1&0&0&0&0&0&0&0&1&1&1&1&0&0&1&1&0&0&1&0\\
	0&1&0&0&0&0&0&0&0&0&0&0&0&1&1&1&1&1&1&1&0&0&1&0&0&0&1&0&0&0&0&0&1&1&1&1\\
0&0&1&0&0&0&0&0&0&0&1&0&0&1&1&0&1&1&1&1&1&0&0&1&1&0&0&0&0&0&1&1&1&1&1&1\\
0&0&0&1&0&0&0&0&0&0&0&0&0&0&1&0&0&1&0&0&1&1&1&0&0&1&1&0&1&1&0&0&0&1&1&1\\
0&0&0&0&1&0&0&0&0&0&0&0&0&1&1&0&1&1&1&1&0&0&0&0&0&0&1&1&0&0&1&1&0&1&0&1\\
0&0&0&0&0&1&0&0&0&0&1&0&0&0&0&0&0&0&1&1&0&1&1&1&1&1&1&1&0&1&0&1&0&1&0&1\\
0&0&0&0&0&0&1&0&0&0&0&0&0&1&1&0&0&1&1&1&0&1&1&1&1&1&1&0&1&0&0&0&1&0&1&1\\
0&0&0&0&0&0&0&1&0&0&1&0&0&1&1&1&0&1&0&1&1&0&0&0&1&0&1&0&1&0&1&0&0&1&1&1\\
0&0&0&0&0&0&0&0&1&0&1&0&0&1&0&0&1&1&1&1&0&0&0&0&1&0&0&0&0&1&0&1&1&1&0&1\\
0&0&0&0&0&0&0&0&0&1&0&0&0&1&1&0&0&1&0&1&1&1&0&0&0&0&1&1&1&0&1&0&0&1&1&0\\
0&0&0&0&0&0&0&0&0&0&0&1&0&1&0&0&0&0&0&0&0&1&0&1&0&0&1&0&1&1&1&0&0&0&0&1\\
0&0&0&0&0&0&0&0&0&0&0&0&1&1&0&0&0&0&0&1&1&0&0&0&0&1&1&0&1&0&0&0&0&1&0&1
		\end{array}\right)$}
	\end{equation*}
	and one can check that, for all $j\in[12]$, we have $t_j^\R(\C)=36$ which means that the only Ravagnani-type anticode that intersects $\C$ in a subspace of dimension at least $j$ is the full space $\F_2^3\otimes\F_2^3\otimes\F_2^4$. In particular, $\C$ is $j$-TMRD for all $j\in[12]$ and, by Lemma~\ref{lem:CCdual1TMRD}, we have that also $\C^\perp$ is $1$-TMRD and $t_1^\R(\C)=24$. Finally, we have $t_1^\R(\C)+t_1^\R(\C^\perp)=36+24$ in line with Proposition~\ref{prop:boundt1+t1perp}.
\end{example}

\begin{example}
	Let $\C$ be the code as in Example~\ref{ex:1-TMRD}. We already noted that $\C$ is not $j$-TMRD for all $j\in[13]$ in line with Proposition~\ref{prop:jTMRD}. Indeed, one can check that $\C^\perp$ is the $\F_2$-$[2\times 3\times 4, 11]$ code whose generator matrix is
	\begin{equation*}
		\left(\begin{array}{*{3}{*{3}c;{3pt/3pt}*{3}c|}*{3}c;{3pt/3pt}*{3}c}
			1&0&0&0&0&0&0&0&0&0&1&1&1&0&0&0&0&0&0&1&0&1&1&1\\
			0&1&0&0&0&0&0&0&0&0&1&0&1&0&1&0&0&0&1&0&1&1&0&0\\
			0&0&1&0&0&0&0&0&0&0&1&1&1&0&1&1&1&1&1&1&0&1&1&1\\
			0&0&0&1&0&0&0&0&0&0&0&1&0&0&0&1&1&1&1&0&0&1&0&0\\
			0&0&0&0&1&0&0&0&0&0&0&0&0&0&0&0&1&0&0&1&1&0&0&0\\
			0&0&0&0&0&1&0&0&0&0&0&1&0&0&0&0&1&0&0&0&0&0&0&0\\
			0&0&0&0&0&0&1&0&0&0&0&0&0&0&0&0&1&0&0&0&1&1&0&0\\
			0&0&0&0&0&0&0&1&0&0&1&1&0&0&1&1&1&0&0&1&1&1&0&0\\
			0&0&0&0&0&0&0&0&1&0&0&1&0&0&0&1&0&0&1&1&1&0&0&1\\
			0&0&0&0&0&0&0&0&0&1&0&0&1&0&0&1&1&0&0&1&0&1&1&1\\
			0&0&0&0&0&0&0&0&0&0&0&0&0&1&0&1&0&0&0&1&1&0&1&0
		\end{array}\right)
	\end{equation*}
	and $\C^\perp$ is not $1$-TMRD since
	\begin{equation*}
		t_1^\R(\C^\perp)=12\neq 24=24-12\left\lfloor\frac{11-1}{12}\right\rfloor.
	\end{equation*}
	Finally, we have $t_1^\R(\C)+t_1^\R(\C^\perp)=12+12<24+12$ in line with Proposition~\ref{prop:boundt1+t1perp}.
\end{example}

The following results extend~\cite[Lemmas~4.12, 4.13, 4.14]{byrne2020rank} respectively. We omit the proofs of the first two results as they are similar to the ones of ~\cite[Lemmas~4.12 and 4.13]{byrne2020rank}.

	\begin{lemma}
   	\label{lem:oldpaper1}
   		Let $j\in\{2,\ldots,k-1\}$. Write $j=k+\rho-\alpha\frac{n}{n_1}$ for $\rho\in\{1,\ldots,\frac{n}{n_1}\}$ and  $\alpha\in\left\{1,\ldots, \left\lfloor\frac{n_1}{n}(k+\rho-1)\right\rfloor\right\}$. If the code $\C$ is $j$-TBMD with respect to $\A^\R$ and $r\leq k^\perp$ then for all $\gamma\in\{1,\ldots,\alpha\}$ we have $t_{k+\rho-\gamma\frac{n}{n_1}}=n-\frac{n}{n_1}(\gamma-1)$.
   	\end{lemma}
   	
   	\begin{lemma}
   	\label{lem:oldpaper2}
   		Let $j\in\{2,\ldots,k-1\}$, $\rho\in\left\{2,\ldots,\frac{n}{n_1}\right\}$ and $k^\perp\in\{1,\ldots,\rho\}$. If $\C$ is $j$-TBMD then one of the following holds.
   		\begin{enumerate}[label=(\arabic*)]
   			\item $j>k+1-\frac{n}{n_1}$ and $t_{k+1-\gamma\frac{n}{n_1}}^\R=n-\frac{n}{n_1}\gamma$ for all $\gamma\in\left\{1,\ldots, \left\lfloor\frac{n_1}{n}k\right\rfloor\right\}$.
   			\item There exists an integer $\alpha\in\left\{1,\ldots, \left\lfloor\frac{n_1}{n}k\right\rfloor\right\}$ such that $k+1-(\alpha+1)\frac{n}{n_1}<j\leq k+1-\alpha\frac{n}{n_1}$, $\left(t_1^\R\right)^\perp=\frac{n}{n_1}(\alpha+1)$ and 
   				\begin{equation*}
   					\begin{cases}
   						\displaystyle t_{k+1-\gamma\frac{n}{n_1}}^\R =n-\frac{n}{n_1}(\gamma-1) & \textup{ if } \gamma\in\{1,\ldots,\alpha\},\\[12pt]
   						\displaystyle t_{k+1-\gamma\frac{n}{n_1}}^\R =n-\frac{n}{n_1}\gamma & \textup{ if } \gamma\in\left\{\alpha+1,\ldots,\left\lfloor\frac{n_1}{n}k\right\rfloor\right\}.
   					\end{cases}
   				\end{equation*}
   		\end{enumerate}
   	\end{lemma}
   	
   	\begin{lemma}
   	\label{lem:tkR}
   		If $\C$ is $k$-TBMD with respect to $\A^\R$ then $t_k^\R=n$.
   	\end{lemma}
   	\begin{proof}
   		Suppose toward a contradiction that there exists a code $\C$ such that $\C$ is $k$-TBMD with respect to $\A^\R$ and $t_k^\R=\delta<n$. Note that $\delta=\frac{n}{n_1}\gamma$ for some $\gamma\in[n-1]$, by the definition of generalized tensor weights with respect to $\A^\R$. We have
   		\begin{equation*}
   			\delta=\left\{\dimq{A}:A\in\A^\R\textup{ and }\dimq{\C\cap A}\geq k\right\}=\left\{\dimq{A}:A\in\A^\R\textup{ and }\C\leq A\right\}.
   		\end{equation*}
   		Therefore, there must exist a Ravagnani-type anticode $A$ containing $\C$. If $\C\leq A$ then $A^\perp\leq\C^\perp$ and $A^\perp\in\A^\R$ since the set of Ravagnani-type anticodes is closed under duality by Proposition~\ref{prop:charoverlineA}. The code $\C^\perp$ must contain a simple tensor of since $A^\perp$ is perfect and so $\left(t_1^\R\right)^\perp=\frac{n}{n_1}$. This implies $n-t_k^\R-\left(t_1^\R\right)^\perp\geq n-\delta-1\geq 0$ and we get a contradiction.
   	\end{proof}

   	The next result generalizes~\cite[Theorem~4.15]{byrne2020rank}.
   	
   	\begin{theorem}
   	\label{thm:jTBMDjTMRD}
   		If $\C$ is $j$-TBMD with respect to $\A^\R$ then $\C$ is $j$-TMRD with respect to $\A^\R$.
   	\end{theorem}
   	\begin{proof}
   		We already observed that the trivial codes $\{0\}$ and $\F$ are both $1$-TMRD with respect $\A^\R$. Note that 
   		\begin{equation*}
   			n-t_1^\R(\{0\})-t_1^\R(\F)=n-n-\frac{n}{n_1}-\frac{n}{n_1}=-2\frac{n}{n_1}<0
   		\end{equation*} 
   		which implies that $\{0\}$ and $\F$ are also $1$-TBMD with respect $\A^\R$. Lemma~\ref{lem:tkR} implies $t_k(\C)=n$ if $\C$ is $1$-TBMD with respect $\A^\R$. Hence, we have $n=n-\frac{n}{n_1}\left\lfloor\frac{n_1}{n}(k-k)\right\rfloor$ and $\C$ is $1$-TMRD with respect $\A^\R$. If $\C$ is $1$-TBMD then $n<t_1^\R+\left(t_1^\R\right)^\perp$ and, by~\cite[Proposition~5.10]{byrne2021tensor} and Proposition~\ref{prop:boundt1+t1perp}, we get 
   		\begin{equation*}
   			t_1^\R+\left(t_1^\R\right)^\perp=\begin{cases}
   				n+\frac{n}{n_1} & \textup{ if } \frac{n}{n_1}\nmid k,\\
   				n+2\frac{n}{n_1} & \textup{ if } \frac{n}{n_1}\mid k.
   			\end{cases}
   		\end{equation*}
   		and $\C$ is $1$-TMRD with respect to $\A^\R$. The remainder of the proof is similar to the proof of~\cite[Theorem~4.15]{byrne2020rank} and follows by Lemmas~\ref{lem:oldpaper1} and~\ref{lem:oldpaper2}.
   	\end{proof}
   	
   The following generalizes~\cite[Theorem~4.19]{byrne2020rank}. We omit the proof as it is similar to the one of~\cite[Theorem~4.19]{byrne2020rank}.
   	
   	\begin{theorem}
   		Let $j\in\{2,\ldots,k\}$. If $\C$ is minimally $j$-TBMD with respect to $\A^\R$ then $\C$ is not $(j-1)$-TMRD with respect $\A^\R$.
   	\end{theorem}

The following example shows that the converse of Theorem~\ref{thm:jTBMDjTMRD} does not hold.

\begin{example}
	Let $\C$ be the code in Example~\ref{ex:1-TMRD} and we already observed that $\C$ is $1$-TMRD but not $2$-TMRD with respect to $\A^\R$. On the other hand, we have that $\C$ is not $1$-TBMD with respect to $\A^\R$ since  $24-t_1^\R(\C)-t_1^\R(\C^\perp)=24-12-12=0\not<0$. We can also notice that $\C$ is minimally $3$-TBMD with respect to $\A^\R$ since 
			\begin{equation*}
				24-t_2^\R(\C)-t_1^\R(\C^\perp)=24-12-12\not<0\quad \textup{ and }\quad 24-t_3^\R(\C)-t_1^\R(\C^\perp)=24-24-12<0.
			\end{equation*}
		and $3$-TMRD with respect to $\A^\R$ since
		\begin{equation*}
			t_3^\R(\C)=24=24-12\left\lfloor\frac{13-3}{12}\right\rfloor.
		\end{equation*}
		Finally, in Example~\ref{ex:1-TMRD}, we showed that $\C$ is not $2$-TMRD with respect to $\A^\R$. This is in line with Theorem~\ref{thm:jTBMDjTMRD}.
\end{example}

In~\cite{byrne2020rank}, it was observed that the property of being minimally $j$-TBMD does not obey to a duality statement in general. In the following example we show that there exists codes with the same dimension whose duals are not $j$-TBMD with respect to $\A^\R$ for the same $s\in[n-k]$.

\begin{example}
	Let $\C_1,\C_2$ be the $\F_2$-$[3\times 3\times 4,11]$ codes generated by 
	\begin{equation*}
		\resizebox{\textwidth}{!}{$
		\left(\begin{array}{*{3}{*{3}c;{3pt/3pt}*{3}c;{3pt/3pt}*{3}c|}*{3}c;{3pt/3pt}*{3}c;{3pt/3pt}*{3}c}
1&0&0&0&0&0&0&0&0&0&0&0&1&1&1&0&0&1&0&0&0&0&1&0&0&0&1&0&1&0&0&1&0&0&0&1\\
0&1&0&0&0&0&0&0&0&0&0&1&1&1&1&0&1&0&0&0&1&1&1&0&1&0&0&1&0&1&0&1&1&1&0&0\\
0&0&1&0&0&0&0&0&0&0&0&0&1&1&1&0&1&1&0&1&0&1&0&1&0&1&0&1&0&1&0&1&1&1&0&0\\
0&0&0&1&0&0&0&0&0&0&0&0&0&0&1&0&1&0&0&1&1&0&0&1&1&0&1&1&1&1&1&1&1&1&1&1\\
0&0&0&0&1&0&0&0&0&0&0&0&0&1&0&0&0&0&0&0&1&1&1&1&1&0&1&1&1&1&1&0&0&0&0&0\\
0&0&0&0&0&1&0&0&0&0&0&1&1&1&0&1&0&1&0&1&0&1&0&1&1&0&0&0&0&0&1&1&1&0&0&0\\
0&0&0&0&0&0&1&0&0&0&0&0&1&1&0&0&0&0&0&1&0&1&1&1&0&1&1&1&1&1&1&0&0&1&0&0\\
0&0&0&0&0&0&0&1&0&0&0&0&1&0&0&1&0&1&1&1&0&1&0&1&1&1&0&1&0&1&1&0&0&0&0&1\\
0&0&0&0&0&0&0&0&1&0&0&0&1&1&0&1&1&1&1&1&0&1&0&1&1&0&0&1&0&1&0&1&1&1&1&0\\
0&0&0&0&0&0&0&0&0&1&0&0&1&1&0&0&0&0&1&0&0&0&0&0&0&1&0&0&0&0&0&1&0&0&0&0\\
0&0&0&0&0&0&0&0&0&0&1&1&0&0&1&1&0&1&1&1&0&1&1&0&1&1&1&1&0&1&0&0&1&1&0&1	
		\end{array}\right)$}
	\end{equation*}
	and
	\begin{equation*}
		\resizebox{\textwidth}{!}{$
		\left(\begin{array}{*{3}{*{3}c;{3pt/3pt}*{3}c;{3pt/3pt}*{3}c|}*{3}c;{3pt/3pt}*{3}c;{3pt/3pt}*{3}c}
1&0&0&0&0&0&0&0&0&0&0&1&0&0&1&0&1&1&0&0&0&0&1&0&0&0&0&1&1&0&1&0&1&1&0&1\\
0&1&0&0&0&0&0&0&0&0&0&0&0&1&0&0&1&1&1&0&0&0&0&0&0&1&0&0&1&0&1&0&1&0&1&1\\
0&0&1&0&0&0&0&0&0&0&0&1&1&1&1&1&1&0&0&0&1&0&1&1&1&1&0&0&1&0&1&0&0&1&0&0\\
0&0&0&1&0&0&0&0&0&0&0&1&0&1&1&0&0&0&1&0&0&1&1&0&1&1&1&1&1&0&0&0&1&1&0&1\\
0&0&0&0&1&0&0&0&0&0&0&1&0&1&1&0&0&0&0&1&1&1&1&0&1&0&1&0&0&0&1&1&0&0&1&1\\
0&0&0&0&0&1&0&0&0&0&0&1&1&1&0&0&1&0&0&0&1&0&1&1&0&0&1&1&0&1&0&1&0&1&0&0\\
0&0&0&0&0&0&1&0&0&0&0&1&1&0&0&0&0&1&0&0&0&0&0&1&1&0&1&0&0&1&0&1&1&1&1&0\\
0&0&0&0&0&0&0&1&0&0&0&1&0&0&1&0&0&1&1&0&0&1&1&0&1&1&0&0&0&1&0&0&0&1&1&0\\
0&0&0&0&0&0&0&0&1&0&0&0&1&0&0&1&1&1&1&1&0&1&0&0&1&1&1&0&0&1&0&0&0&0&1&1\\
0&0&0&0&0&0&0&0&0&1&0&0&0&1&0&1&1&0&1&0&0&1&1&0&1&0&0&0&0&0&1&1&0&0&1&0\\
0&0&0&0&0&0&0&0&0&0&1&0&0&1&0&1&1&0&0&0&1&1&1&1&0&0&1&0&1&0&0&1&0&1&0&1	
		\end{array}\right)$}
	\end{equation*}
	respectively. One can check the following.
	\begin{itemize}
	\setlength\itemsep{4pt}
		\item The generalized tensor weights of $\C_1$ with respect to $\A^\R$ are
		\begin{equation*}
			t_1^\R(\C_1)=12,\qquad t_2^\R(\C_1)=24,\qquad t_3^\R(\C_1)=\cdots=t_{11}^\R(\C_1)=36.
		\end{equation*}
		\item The generalized tensor weights of  $\C_2$ with respect to $\A^\R$ are
		\begin{equation*}
			t_1^\R(\C_2)=t_2^\R(\C_2)=24,\qquad t_3^\R(\C_2)=t_{11}^\R(\C_2)=36.
		\end{equation*}
		\item The generalized tensor weights of  $\C_1^\perp$ with respect to $\A^\R$ are
		\begin{align*}
			&t_1^\R(\C_1^\perp)=\cdots=t_3^\R(\C_1^\perp)=12,\qquad t_4^\R(\C_1^\perp)=\cdots=t_{14}^\R(\C_1^\perp)=24,\qquad \\
			&t_{15}^\R(\C_1^\perp)=\cdots=t_{25}^\R(\C_1^\perp)=36
		\end{align*}
		\item The generalized tensor weights of  $\C_2^\perp$ with respect to $\A^\R$ are
		\begin{align*}
			&t_1^\R(\C_2^\perp)=\cdots=t_3^\R(\C_2^\perp)=12,\qquad  t_4^\R(\C_1^\perp)=\cdots=t_{13}^\R(\C_1^\perp)=24,\\ &t_{14}^\R(\C_1^\perp)=\cdots=t_{25}^\R(\C_1^\perp)=36.
		\end{align*}
	\end{itemize}
	Moreover, the following hold
	\begin{itemize}
	\setlength\itemsep{4pt}
		\item $36-t_2^\R(\C_1)-t_1(\C_1^\perp)=36-t_2^\R(\C_2)-t_1(\C_2^\perp)=36-24-12=0\not<0$.
		\item $36-t_3^\R(\C_1)-t_1(\C_1^\perp)=36-t_3^\R(\C_2)-t_1(\C_2^\perp)=36-36-12=-12<0$.
		\item $36-t_{14}^\R(\C_1^\perp)-t_1(\C_1)=36-24-12=0\not<0$.
		\item $36-t_{15}^\R(\C_1^\perp)-t_1(\C_1)=36-36-12=-12<0$.
		\item $36-t_{3}^\R(\C_2^\perp)-t_1(\C_2)=36-12-24=0\not<0$.
		\item $36-t_{15}^\R(\C_1^\perp)-t_1(\C_1)=36-24-24=-12<0$.
	\end{itemize}
	This implies that $\C_1$ and $\C_2$ are both minimally $3$-TBMD, $\C_1^\perp$ is $15$-TBMD and $\C_2^\perp$ is $4$-TBMD.
\end{example}

	
\section{The Tensor Zeta Function}
\label{sec:zeta}
	Following the work in~\cite{blanco2018rank} and~\cite{byrne2020rank}, in this section we introduce the zeta function for generalized tensor weights related to the different collections of anticodes classified in~\cite{byrne2021tensor} and we then specialize the theory for the Delsarte and Ravagnani-type anticodes. Throughout this and the next section, we consider polynomials that are members of the polynomial rings $\Q[T]$ and $\Q[X,Y,T]$. We introduce the following notation.
	
\begin{notation}
\label{not:delta}
	We let $\delta_i:=\{s\in [r]:n_i=n_s\}$, for any $i\in [r]$, and 
	\begin{equation*}
		\Delta:=\begin{cases}
			\{n_1,\ldots, n_r\} & \textup{ if } |\delta_{r}|> 1,\\[3pt]
			\{n_1,\ldots, n_{r-1}\} & \textup{ if } |\delta_{r}|= 1.
		\end{cases}
	\end{equation*} 
\end{notation}

Observe that $|\delta_{i}|\geq 1$ for all $i\in[r]$. We define the following.

\begin{definition}
\label{def:ba}
	For any $a\in\Z$, the $(a,j)$-\textbf{normalized tensor binomial moment} of $\C$ with respect to $\A$ is 
	\begin{equation*}
		b_a^{(j)}(\C):=\begin{cases}
			0 & \textup{ if } a<0,\\[3pt]
			\displaystyle\frac{B_{a+t_j}^{(j)}(\C)}{|\A_{a+t_j}|} & \textup{ if } 0\leq a\leq n-t_j-s_1^\perp,\\[12pt]
			\qbin{k+a+t_j-n}{j} & \textup{ if } a>n-t_j-s_1^\perp \textup{ and } \mathcal{P}_\A,\\[15pt]
			0 &\textup{ if }  a>n-t_j-s_1^\perp \textup{ and } \neg\mathcal{P}_\A,
		\end{cases}
	\end{equation*}
	where $\mathcal{P}_\A$ is the predicate defined as follows
	\begin{equation*}
		\mathcal{P}_\A:=\begin{cases}
			\top & \textup{ if } \A=\A^\ps,\\[2pt]
			\A_a^\cl\neq \emptyset & \textup{ if } \A=\A^\cl \textup{ and } a\in\{0,\ldots n\},\\[2pt]
			a\in\{a_1\cdots a_r:a_i\in\Z_{\geq n_i} \textup{ for all } i\in[r] \}  & \textup{ if } \A=\A^\cl \textup{ and } a> n,\\[2pt]
			n\mid (a+t_j^\D)m \textup{ for some } m\in\Delta & \textup{ if } \A=\A^\D,\\[2pt]
			n\mid (a+t_j^\R)n_1 & \textup{ if } \A=\A^\R.
		\end{cases}
	\end{equation*}
\end{definition}

Observe that, for any $a\in\Z$, the condition $n\mid (a+t_j^\R)n_1$ is equivalent to $n\mid an_1$ since $n\mid t_j^\R n_1$ by definition of Ravagnani-type anticodes. We now explicitly determine the cardinality of the set $\A_a^\D$ and $\A_a^\R$ for all $a\in \{0,\ldots n\}$ using the fact that $\A^\D$ and $\A^\R$ are product lattices.

\begin{proposition}
\label{prop:carAD}
For any $a\in\{0,\ldots n\}$ we have
	\begin{equation*}
	\left|\A_a^\D\right|=
	\begin{cases}
		0 & \textup{ if } n \nmid am \textup{ for all } m\in \Delta,\\
    	\displaystyle \sum_{\substack{m\in \Delta\\ n\mid am}}\qbin{m}{a\frac{m}{n}} & \textup{ if } n\mid am \textup{ for some } m\in \Delta \textup{ and } a\notin\{0,n\},\\
  		1 & \textup{ if } a\in\{0,n\}.
     \end{cases}
	\end{equation*}
Moreover, if $n \nmid am$ for all $m\in\Delta$ then $B_a^{(\D,j)}(\C)=0$ and $W_a^{(\D,j)}(\C)=0$.
\end{proposition}
\begin{proof}
Clearly, $\A^\D_0=\{\<0\>_{\Fq}\}$ and $\A^\D_n=\{\F\}$, and we can assume $a\in[n-1]$ in the remainder of the proof. By Definition~\ref{def:anticodes}, we have that $A\in\A^\D$ if and only if 
	\begin{equation*}
		A=\bigotimes_{s=1}^{i-1}\Fq^{n_s}\otimes A^{(i)}\otimes\bigotimes_{s=i+1}^{r}\Fq^{n_s}
	\end{equation*}
	for some $A^{(i)}\leq \Fq^{n_i}$ with $n_i\in \Delta$. It is not hard to check that a Delsarte-type anticode of dimension $a$ exists if and only if $n\mid am$ for some $m\in \Delta$. Therefore, if $n\nmid am$ for all $m\in \Delta$ then $B_a^{(\D,j)}(\C)=0$ and $W_a^{(\D,j)}(\C)=0$. It remains to prove the case $n\mid am$ for some $m\in \Delta$. Define the set 
	\begin{equation*}
   		\A_{a,i}^\D:=\left\{\bigotimes_{j=1}^{i-1}\Fq^{n_j}\otimes A^{(i)}\otimes\bigotimes_{j=i+1}^{r}\Fq^{n_j}:A^{(i)}\leq\Fq^{n_i} \textup{ and } \dimq{A^{(i)}}=a\right\}
	\end{equation*}
	and observe that if $n_i=n_s$ then $\left|\A_{a,i}^\D\right|=\left|\A_{a,s}^\D\right|$, for all $i,s\in[r]$. Finally, if there exists an $m\in N$ such that $n\mid am$ then we have
	\begin{align*}
		\left|\A_a^\D\right|=\left|\bigcupdot_{i=1}^r\A_{a,i}^\D\right|=\sum_{i=1}^r\left|\A_{a,i}^\D\right|=\sum_{\substack{i=1\\n\mid ai}}^r\left|\A_{a,i}^\D\right|=\sum_{\substack{m\in \Delta\\ n\mid am}}\qbin{m}{a\frac{m}{n}}
	\end{align*}
	which concludes the proof.
\end{proof}

A similar argument implies the following result. We include a proof for completeness.

	\begin{proposition}
    \label{prop:carAR}
        For any $a\in\{0,\ldots n\}$ we have
        \begin{equation*}
            \left|\A_a^\R\right|=
            \begin{cases}
                0 & \textup{ if } n \nmid an_1,\\
               \displaystyle |\delta_1|\qbin{n_1}{a\frac{n_1}{n}} & \textup{ if } n\mid an_1 \textup{ and } a\notin\{0,n\},\\
  				1 & \textup{ if } a\in\{0,n\}.
            \end{cases}
        \end{equation*}
        Moreover, if $n \nmid an_1$ then $B_a^{(\R,j)}(\C)=0$ and $W_a^{(\R,j)}(\C)=0$.
    \end{proposition}
    \begin{proof}
    	Clearly, $\A_0^{\R}=\{\<0\>_{\Fq}\}$ and $\A_n^{\R}=\{\F\}$, and we assume $a\in[n-1]$ in the remainder of the proof. By~\cite[Theorem~4.21]{byrne2021tensor}, we have that $A\in\A^\R$ if and only if 
   		\begin{equation*}
   			A=\bigotimes_{j=1}^{i-1}\Fq^{n_j}\otimes A^{(i)}\otimes\bigotimes_{j=i+1}^{r}\Fq^{n_j}
   		\end{equation*}
   		for some $A^{(i)}\leq \Fq^{n_i}$ such that $n_i=n_1$. Therefore, one can easily check that if $n\nmid an_1$ then it cannot exist an $A\in\A^\R$ of dimension $a$ and therefore $B_a^{(\R,j)}(\C)=0$ and $W_a^{(\R,j)}(\C)=0$. Finally, if $n\mid an_1$ then we have
    	\begin{align*}
    		\left|\A_a^\R\right|&=\left|\bigcupdot_{\substack{i=1\\[2pt]n_i=n_1
    		}}^r \left\{\bigotimes_{s=1}^{i-1}\Fq^{n_s}\otimes A^{(i)}\otimes\bigotimes_{s=i+1}^{r}\Fq^{n_s}:A^{(i)}\leq\Fq^{n_i}\textup{ and }\dimq{A^{(i)}}=a\frac{n_1}{n}\right\}\right|\\
    		&=|\delta_1|\left|\left\{ A^{(1)}\otimes\bigotimes_{s=2}^{r}\Fq^{n_s}:A^{(1)}\leq\Fq^{n_1}\textup{ and }\dimq{A^{(1)}}=a\frac{n_1}{n}\right\}\right|
    	\end{align*}
    	which concludes the proof.
   	\end{proof}

\begin{remark}
	Observe that for $r=2$ and $\A=\A^\R$ we recover the normalized generalized binomial moments given in~\cite[Definition~3.10]{byrne2020rank}. More in detail, denote by $\hat b_a^{(j)}(\C)$ the quantities defined in~\cite[Definition~3.10]{byrne2020rank} and we have $b_{an_2}^{(\D,j)}(\C)=b_{an_2}^{(\R,j)}(\C)=\hat b_a^{(j)}(\C)$ for all $a\in\Z$.
\end{remark}

We conclude this section by introducing the generalized zeta function of a tensor code as the generating function of the normalized generalized tensor binomial moments. 

\begin{definition}
	 The $j$\textbf{-th tensor zeta function} of $\C$ is 
       \begin{equation*}
           Z_{\C}^{(j)}(T):=\sum_{a\in\Z}b_a^{(j)}(\C)T^a=\sum_{a\geq 0}b_a^{(j)}(\C)T^a.
       \end{equation*}
\end{definition}

It is interesting to observe that if a code is $j$-TBMD then its normalized generalized tensor binomial moments and its generalized tensor zeta function indexed by $j$ are determined by the code parameters. The following result is immediate.

	\begin{proposition}
	\label{prop:ZTBMD}
		Let $\C$ be $j$-TBMD. The following hold for all $s\in\{j,\ldots,k\}$.
		\begin{enumerate}[label=(\arabic*)]
			\item For any $a\in\Z_{\geq 0}$, we have $b_a^{(s)}(\C)=
			\begin{cases}
				\qbin{k+a-t_s-n}{s} & \textup{ if } \mathcal{P}_\A,\\[15pt]
				0 & \textup{ if } \neg\mathcal{P}_\A,
			\end{cases}$
			
			\item $\displaystyle Z_\C^{(s)}(T)=\sum_{\substack{a\geq 0\\[1pt]\mathcal{P}_\A}}\qbin{k+a-t_s-n}{s}T^a$.
		\end{enumerate}
	\end{proposition}

\section{The Tensor Zeta Function for Ravagnani-type Anticodes}   
\label{sec:zetaR}
    Inspired by the theory developed in~\cite{byrne2020rank}, in this section we extend the study of the zeta function for tensor weights related to Ravagnani-type anticodes. We now recall the definition of the $q$-\textbf{Bernstein polynomial}. The reader is referred to~\cite{lupas1987q} for further details.
    
    \begin{definition}
        Let $a,b$ be nonnegative integers such that $a\leq b$. The $(b,a)$-\textbf{th} $q$-\textbf{Bernstein polynomial} is
        \begin{equation*}
            \B_{b,a}(X,Y;q):=\qbin{b}{a}Y^a\prod_{c=0}^{b-a-1}(X-q^cY).
        \end{equation*}
    \end{definition}
    
     We also recall that the sets $\{X^{b-a}Y^{a}:a,b\in\Z_{\geq 0}, a\leq b\}$ and $\{\B_{b,a}(X,Y;q):a,b\in\Z_{\geq 0}, a\leq b\}$ are $\Q$-bases for the ring of homogeneous polynomials of degree $b$. The inversion formula associated with these polynomials is
    \begin{equation*}
        X^{b-a}Y^a=\qbin{b}{a}^{-1}\sum_{c=a}^b\qbin{c}{a}\B_{b,c}(X,Y;q).
    \end{equation*}

A direct application of Proposition~\ref{prop:carAR} to Proposition~\ref{prop:oldBaWb} gives the following result.

\begin{proposition}
	\label{prop:BaWb}
	The following hold for all $a,b\in\{0,\ldots,n\}$.
	\begin{enumerate}[label=(\arabic*)]
		\item $\displaystyle B_a^{(\R,j)}(\C)=\sum_{b=0}^aW_b^{(\R,j)}(\C)|\delta_{1}|\qbin{n_1-a\frac{n_1}{n}}{(a-b)\frac{n_1}{n}}$.
		\item $\displaystyle W_b^{(\R,j)}(\C)=\sum_{a=0}^bB_a^{(\R,j)}(\C)|\delta_{1}|\qbin{n_1-b\frac{n_1}{n}}{(b-a)\frac{n_1}{n}}(-1)^{(b-a)\frac{n_1}{n}}q^{\binom{(b-a)\frac{n_1}{n}}{2}}$.
	\end{enumerate}
\end{proposition} 	
\begin{proof}
	Note that by Proposition~\ref{prop:carAR} we have
	\begin{equation*}
            \left|\left\{(A',A)\in\A_b\times\A_a:A'\leq A\right\}\right|=
            \begin{cases}
                0 &\textup{ if }\; n\nmid n_1b \;\textup{ or }\; n\nmid n_1a,\\[2ex]
                \displaystyle |\delta_{1}|\qbin{n_1-b\frac{n_1}{n}}{(a-b)\frac{n_1}{n}} & \textup{ if }\; n\mid n_1b \;\textup{ and }\; n\mid n_1a.
            \end{cases}
        \end{equation*}
    Finally, one can check that the lattices
    \begin{equation*}
    	\left\{A^{(1)}\otimes\bigotimes_{s=2}^{r}\Fq^{n_s}:A^{(1)}\leq\Fq^{n_1}\right\}\qquad \textup{ and }\qquad \left\{A\leq\Fq^{n_1}\right\}
    \end{equation*}
    are isomorphic and so the statement follows by Proposition~\ref{prop:oldBaWb}.
\end{proof}

In the next theorem, we compute the coefficients of $\W_\C^{(\R,j)}(X,Y)$ in terms of the basis given by $q$-Bernstein polynomials. This result can be seen a generalization~\cite[Theorem~3.11]{byrne2020rank} for $r\geq 2$.

\begin{theorem}
   	\label{thm:Wbern}
   		The following holds.
   		\begin{equation*}
   			\W_\C^{(\R,j)}(X,Y)=\sum_{a=t_j^\R\frac{n_1}{n}}^{n_1}b_{a\frac{n}{n_1}-t_j^\R}^{(\R,j)}(\C)\B_{{n_1},a}\left(X^{\frac{n}{n_1}},Y^{\frac{n}{n_1}};q\right).
   		\end{equation*}
   	\end{theorem}
   	\begin{proof}
   		Proposition~\ref{prop:carAR} and the inversion formula for $q$-Bernstein polynomial imply
   		\begin{align*}
   			\W_\C^{(\R,j)}(X,Y)&=\sum_{\tiny\begin{matrix}b\leq n\\ n\mid bn_1\end{matrix}}W_b^{(\R,j)}(\C)X^{n-b}Y^b\\
   			&=\sum_{b=0}^{n_1}W_{b\frac{n}{n_1}}^{(\R,j)}(\C)\left(X^{\frac{n}{n_1}}\right)^{n_1-b}\left(Y^{\frac{n}{n_1}}\right)^b\\
   			&=\sum_{b=t_j^\R\frac{n_1}{n}}^{n_1}W_{b\frac{n}{n_1}}^{(\R,j)}(\C)\qbin{n_1}{b}^{-1}\sum_{a=b}^{n_1}\qbin{a}{b}\B_{{n_1},a}\left(X^{\frac{n}{n_1}},Y^{\frac{n}{n_1}};q\right)\\ &=\sum_{a=t_j^\R\frac{n_1}{n}}^{n_1}B_{n_1,a}\left(X^{\frac{n}{n_1}},Y^{\frac{n}{n_1}};q\right)\qbin{n_1}{a}^{-1}\sum_{b=t_j^\R\frac{n_1}{n}}^a\qbin{n_1-b}{a-b}W_{b\frac{n}{n_1}}^{(\R,j)}(\C)
   		\end{align*}
   		where the last equality follows by Lemma~\ref{lem:bin}. Finally, by Proposition~\ref{prop:BaWb} and Definition~\ref{def:ba}, we have
        \begin{align*}
        	\W_\C^{(\R,j)}(X,Y)&=\sum_{a=t_j^\R\frac{n_1}{n}}^{n_1}\B_{{n_1},a\frac{n_1}{n}}\left(X^{\frac{n}{n_1}},Y^{\frac{n}{n_1}};q\right)\qbin{n_1}{a}^{-1}\sum_{b=t_j^\R\frac{n_1}{n}}^a\qbin{n_1-a}{a-b}W_{b\frac{n}{n_1}}^{(\R,j)}(\C)\\
        	&=\sum_{a=t_j^\R\frac{n_1}{n}}^{n_1}\B_{{n_1},a}\left(X^{\frac{n}{n_1}},Y^{\frac{n}{n_1}};q\right)\qbin{n_1}{a}^{-1}|\delta_{1}|^{-1}B_{a\frac{n}{n_1}}^{(\R,j)}(\C)\\
        	&=\sum_{a=t_j^\R\frac{n_1}{n}}^{n_1}b_{a\frac{n}{n_1}-t_j^\R}^{(\R,j)}(\C)\B_{{n_1},a}\left(X^{\frac{n}{n_1}},Y^{\frac{n}{n_1}};q\right)
        \end{align*}
        which concludes the proof.
   	\end{proof}

   	The following result provides a connection between the $q$-Bernstein polynomials, the $j$-th generalized tensor zeta function and the $j$-th generalized tensor weight enumerator.
   	
   	\begin{proposition}
   	\label{prop:varphi}
   		Define the function 
   		\begin{equation*}
   			\varphi_a(X,Y,T):=\sum_{b=0}^{a}B_{a,b}(X,Y;q)T^{a-b}.
   		\end{equation*}
   		Then $\W_\C^{(\R,j)}(X,Y)$ is the coefficient of $T^{n-t_j^\R}$ in the expression $Z_\C^{(\R,j)}(T)\varphi_{n_1}\left(X^{\frac{n}{n_1}},Y^{\frac{n}{n_1}},T^{\frac{n}{n_1}}\right)$.
   	\end{proposition}
   	\begin{proof}
   		We have 
   		\begin{align*}
   			Z_\C^{(\R,j)}(T)\varphi_{n_1}&\left(X^{\frac{n}{n_1}},Y^{\frac{n}{n_1}},T^{\frac{n}{n_1}}\right)\\
   			&=\left(\sum_{a\geq 0}b_{a\frac{n}{n_1}}(\C)\left(T^{\frac{n}{n_1}}\right)^a\right)\left(\sum_{b=0}^{n_1}B_{n_1,b}\left(X^{\frac{n}{n_1}},Y^{\frac{n}{n_1}};q\right)\left(T^{\frac{n}{n_1}}\right)^{n_1-b}\right)\\
   			&=\sum_{a\geq 0}\sum_{b=0}^{n_1}b_{a\frac{n}{n_1}}(\C)B_{n_1,b}\left(X^{\frac{n}{n_1}},Y^{\frac{n}{n_1}};q\right)\left(T^{\frac{n}{n_1}}\right)^{n_1-b+a}\\
   			&=\sum_{b=0}^{n_1}\left(T^{\frac{n}{n_1}}\right)^b\sum_{a=0}^bb_{a\frac{n}{n_1}}(\C)B_{n_1,n_1-b+a}\left(X^{\frac{n}{n_1}},Y^{\frac{n}{n_1}};q\right) \mod T^{n+1}\\
   			&=\sum_{b=0}^{n_1}\left(T^{\frac{n}{n_1}}\right)^{n_1-b}\sum_{a=0}^{n_1-b}b_{a\frac{n}{n_1}}B_{n_1,b+a}\left(X^{\frac{n}{n_1}},Y^{\frac{n}{n_1}};q\right) \mod T^{n+1}.
   	\end{align*}
   	Therefore, by Theorem~\ref{thm:Wbern}, $\W_\C^{(\R,j)}(X,Y)$ is the coefficient of $T^{n-t_j^\R}=\left(T^{\frac{n}{n_1}}\right)^{n_1-\frac{n_1}{n}t_j^\R}$ in the expression  $Z_\C^{(\R,j)}(T)\varphi_{n_1}\left(X^{\frac{n}{n_1}},Y^{\frac{n}{n_1}},T^{\frac{n}{n_1}}\right)$.
   	\end{proof}

\begin{example}
	Let $\C$ be the $\F_2$-$[2\times 3\times 4,13]$ code as in Example~\ref{ex:1-TMRD} and recall that $t_2^\R(\C)=12$. One can check the following.
	\begin{align*}
		\varphi_2(X^{12},Y^{12},T^{12})&=X^{24}T^{24}-3X^{12}Y^{12}T^{24} + 3X^{12}Y^{12}T^{12}+2Y^{24}T^{24}-3Y^{24}T^{12}+Y^{24},\\
		Z_\C^{(\R,2)}(T)&=1+ 11180715\, T^{12}+187649967696555\,T^{24}+\cdots\\
		\W_\C^{(\R,2)}(X,Y)&=3X^{12}Y^{12} + 11180712\,Y^{24}.
	\end{align*}
	Therefore, $\W_\C^{(\R,2)}(X,Y)$ is the coefficient of $T^{24-t_2^\R(\C)}=T^{12}$ in the expression
	\begin{align*}
		Z_\C^{(\R,2)}(T)\varphi_2(X^{12},Y^{12},T^{12})=&Y^{24}+(3X^{12}Y^{12} + 11180712\,Y^{24})T^{12}\\
		&+( X^{24} + 33542142\,X^{12}Y^{12} - 33542143\,Y^{24})+\cdots
	\end{align*}
	as predicted by Proposition~\ref{prop:varphi}.
\end{example}

	\begin{theorem}
   	\label{thm:zetapol}
   		There exists a unique polynomial $P_\C^{(\R,j)}(T)\in\Q[T]$ of degree at most  $n_1+j+1-\frac{n_1}{n}(t_j^\R+\left(t_1^\R\right)^\perp)$ such that
   		\begin{equation*}
   			Z_\C^{(\R,j)}(T)=\frac{P_\C^{(\R,j)}\left(T^{\frac{n}{n_1}}\right)}{\displaystyle \prod_{s=0}^j\left(1-q^{\frac{n}{n_1}s}T^{\frac{n}{n_1}}\right)}.
   		\end{equation*}
   		We have that the coefficient of degree $a$ of $P_\C^{(\R,j)}(T)$ is
   		\begin{equation*}
   			p_a^{(\R,j)}(\C)=\sum_{s=0}^{j+1}\bbin{j+1}{s}{\frac{n}{n_1}}(-1)^sq^{\frac{n}{n_1}\binom{s}{2}}b_{\frac{n}{n_1}(a-s)}^{(\R,j)}(\C).
   		\end{equation*}
   	\end{theorem}
   	\begin{proof}
   		Applying Lemma~\ref{lem:bin} to the power series 
   		\begin{equation*}
   			P_\C^{(\R,j)}\left(T^{\frac{n_1}{n}}\right):=Z_\C^{(\R,j)}(T)\prod_{s=0}^j\left(1-q^{\frac{n}{n_1}s}T^{\frac{n_1}{n}}\right)
   		\end{equation*}
   		gives the coefficients $p_a^{(j)}(\C)$ as in the statement. Moreover, a standard computation using Lemma~\ref{lem:bin} shows that 
   		\begin{equation*}
   			\sum_{s=0}^{j+1}\bbin{j+1}{s}{\frac{n}{n_1}}(-1)^sq^{\frac{n}{n_1}\binom{s}{2}}b_{\frac{n}{n_1}(a-s)}^{(j)}(\C)=0
   		\end{equation*}
   		for all $a\notin\{0,\ldots,\lfloor n_1+j+1-\frac{n_1}{n}(t_j^\R+\left(t_1^\R\right)^\perp\rfloor\}$. The statement follows by a change of variable. 
   	\end{proof}
   	
Note that this result extends~\cite[Theorem~5.2]{byrne2020rank} for $r\geq 2$. In the sequel, we refer to the polynomial $P_\C^{(\R,j)}(T)$ as the $j$-\textbf{th tensor zeta polynomial} related to $\A^\R$. As a consequence of Theorem~\ref{thm:jTBMDjTMRD} we have the generalized tensor zeta functions and zeta polynomials of a $j$-TBMD code with respect to $\A^\R$ are partially determined by the code parameters. More in detail, we have the following result which can be seen as a generalization of~\cite[Corollary~5.3]{byrne2020rank}.

	\begin{corollary}
	\label{cor:binomTBMD}
		Write $k=\alpha\frac{n}{n_1}+\rho$ with $\rho\in\{0,\ldots,\frac{n}{n_1}-1\}$. If $\C$ is $j$-TBMD then the following hold for any integer $s\in\{j,\ldots,k\}$.
		\begin{enumerate}[label=(\arabic*)]
			\item $b_a^{(\R,s)}(\C)=\qbin{\rho+a-\frac{n}{n_1}\left\lfloor\frac{n}{n_1}(\rho-s)\right\rfloor}{s}	$ for all $a\geq 0$.
			\item $\displaystyle Z_\C^{(\R,s)}(T)=\sum_{a\geq 0}\qbin{\rho+\frac{n}{n_1}\left(a-\left\lfloor\frac{n}{n_1}(\rho-s)\right\rfloor\right)}{s}\left(T^{\frac{n}{n_1}}\right)^a$.
			\item $\displaystyle p_a^{(\R,s)}(\C)=\sum_{p=0}^{s+1}\bbin{s+1}{p}{\frac{n}{n_1}}(-1)^pq^{\frac{n}{n_1}\binom{p}{2}}\qbin{\rho+\frac{n}{n_1}\left(a-p-\left\lfloor\frac{n}{n_1}(\rho-s)\right\rfloor\right)}{s}$ for all $a\geq 0$.
		\end{enumerate}
	\end{corollary}

Let $\C,\mD\leq \F$ be $j$-TBMD with respect to $\A^\R$ such that $\dimq{\C}\equiv\dimq{\mD}\equiv\rho \mod \frac{n}{n_1}$. Observe that, for any $s\in\{j,\ldots,k\}$, we have $b_a^{(\R,s)}(\C)=b_a^{(\R,s)}(\mD)$, for all $a\geq 0$, and, in particular, they share the same $s$-th generalized tensor zeta function and zeta polynomial.

    \begin{notation}
    \label{not:binombeta}
    	In the sequel, for any $s,a\in\Z_{\geq 0}$ and $\rho\in\{0,\ldots,\frac{n}{n_1}-1\}$ we let
    	\begin{enumerate}[label=(\arabic*)]
    		\item $b_{\rho,a}^{(\R,s)}:=\qbin{\rho+a-\frac{n}{n_1}\left\lfloor\frac{n}{n_1}(\rho-s)\right\rfloor}{s}	$.
    		\item $\displaystyle Z_{\rho}^{(\R,s)}(T):=\sum_{a\geq 0}b_{\rho,a}^{(\R,s)}\left(T^{\frac{n}{n_1}}\right)^a$.
    		\item $\displaystyle p_{\rho,a}^{(\R,s)}:=\sum_{p=0}^{s+1}\bbin{j+1}{p}{\frac{n}{n_1}}(-1)^pq^{\frac{n}{n_1}\binom{p}{2}}\qbin{\rho+\frac{n}{n_1}\left(a-p-\left\lfloor\frac{n}{n_1}(\rho-s)\right\rfloor\right)}{s}$.
    		\item $\displaystyle P_{\rho}^{(\R,s)}(T):=\sum_{a\geq 0}p_{\rho,a}^{(\R,s)}T^a$.
    		\item $\displaystyle\W_{\rho,\ell}^{(\R,j)}(X,Y):=\sum_{a=0}^{n_1-\ell}b_{\rho,a}^{(\R,s)}\B_{n_1,a+\ell}\left(X^{\frac{n}{n_1}},Y^{\frac{n}{n_1}};q\right)$ for all $\ell\in[n_1]$.
    	\end{enumerate}
    \end{notation}	

By Corollary~\ref{cor:binomTBMD}, the objects $b_{\rho,a}^{(\R,s)}, Z_{\rho}^{(\R,s)}(T), p_{\rho,a}^{(\R,s)}, P_{\rho}^{(\R,s)}(T)$ and $\W_{\rho,\ell}^{(\R,s)}(X,Y)$ defined above are those related to a $j$-TBMD code with respect to $\A^\R$ of dimension $k\equiv\rho\mod \frac{n}{n_1}$ and $t_s^\R=\frac{n}{n_1}\ell$, provided that such a code exists. Moreover, it is not difficult to check that, for a given $\rho\in\{0,\ldots,\frac{n}{n_1}-1\}$, the set
  	\begin{equation*}
  		\W_\rho^{(\R,j)}:=\left\{\W_{\rho,\delta}^{(\R,j)}(X,Y):\delta\in[n_1]\right\}
  	\end{equation*}
  	is a $\Q$-basis for the space that contains the $j$-th generalized tensor weight enumerators related to $\A^\R$. We conclude this section by giving a generalization of~\cite[Section~6]{byrne2020rank}, that is computing the coefficients of $\W_\C^{(\R,j)}$ with respect to the basis $\W_\rho^{(\R,j)}$. In particular, we give an explicit expression for the rational numbers $\beta_{\rho,a}^{(\R,j)}$ such that
   	\begin{equation*}
   			\W_\C^{(\R,j)}(X,Y)=\sum_{a=0}^{n_1-\frac{n_1}{n}t_j^\R}\beta_{\rho,a}^{(\R,j)}\W_{\rho,\frac{n_1}{n}t_j^\R+a}^{(\R,j)}(X,Y).
   	\end{equation*}
   	
   	Our approach is similar to the one in~\cite{byrne2020rank} using ordinary\footnote{There are two forms of Bell polynomials, namely the ordinary and the exponential form.} homogeneous Bell polynomials. These objects were introduced in~\cite{bell1927partition} and encode the different ways in which an integer can be partitioned. The reader is refereed to~\cite{comtet2012advanced} and~\cite{port1994polynomial} for further details.  We briefly recall the main definitions and results. Compare the following definition with the characterization of homogeneous exponential partial Bell polynomials given in~\cite[Section~3.3, Theorem~A]{comtet2012advanced}.
   	
   	\begin{definition}
   		Let $a,b\in\Z_{\geq 0}$. The $(a,b)$-\textbf{th homogeneous ordinary partial Bell polynomial} is 
   		\begin{align*}
   			\PP_{a,b}(x_0,x_1,\ldots,x_{a-b+1})&=\sum_{\substack{c_1+c_2+\cdots+c_{a-b+1}=b\\ c_1+2c_2+\cdots+(a-b+1)c_{a-b+1}=a}}\frac{a!}{c_1!c_2!\cdots c_{a-b+1}!}x_0^{a-b}x_1^{c_1}x_2^{c_2}\cdots x_{a-b+1}^{a-b+1}.
   		\end{align*}
   		Moreover, we let $\PP_{0,0}:=1$, $\PP_{a,0}:=0$ and $\PP_{0,b}:=0$. 
   	\end{definition}
   	
   	Observe that the polynomial $\PP_{a,b}(x_0,x_1,\ldots,x_{a-b+1})$ describes the compositions of the integer $a$ into $b$ summands. We recall that the compositions of an integer $a$ are simply ordered partitions, that is the possible ways the integer $a$ can be partitioned where the order matters.
   	
   	\begin{definition}
   		Let $a\in\Z_{\geq 0}$. The The $a$-\textbf{th} (\textbf{homogeneous ordinary}) \textbf{Bell polynomial} is the sum over $b\in\Z_{\geq 0}$ of the $(a,b)$-th homogeneous ordinary partial Bell polynomials, that is
   		\begin{equation*}
   			\PP_{a}(x_0,x_1,\ldots,x_{a})=\sum_{b=0}^a\PP_{a,b}(x_0,x_1,\ldots,x_{a-b+1}).
   		\end{equation*} 
   	\end{definition}
   	
   	We recall an application of the Fa\`a di Bruno formula~\cite{faa1855sullo}. For our purpose, we are interested in a combinatorial formulation of such result involving Bell polynomials. 
   	
   	\begin{lemma}
   		If $(c_a)_{a\in\Z_\geq 0}$ is a sequence and $Q(T)$ is the formal power series 
   		\begin{equation*}
   			Q(T)=1-\sum_{a\geq 0}\frac{c_a}{c_0}T^a
   		\end{equation*}
   		than its multiplicative inverse $R(T)=\frac{1}{Q(T)}$ can be written as 
   		\begin{equation*}
   			R(T)=1+\sum_{a\geq 1}d_aT^a\qquad \textup{ where }\qquad d_a=\left(\frac{1}{c_0}\right)^a\PP_a(c_0,c_1,c_2,\ldots,c_a).
   		\end{equation*}
   	\end{lemma}

   	The next connection between generalized tensor zeta functions and zeta polynomials generalizes~\cite[Lemma~5.8]{byrne2020rank}.
   	
   	\begin{lemma}
   		For all $\rho\in\left\{0,\ldots,\frac{n}{n_1}-1\right\}$ we have 
   		\begin{equation*}
   			Z_\C^{(\R,j)}(T)P_\rho^{(\R,j)}\left(T^\frac{n}{n_1}\right)=Z_\rho^{(\R,j)}(T)P_\C^{(\R,j)}\left(T^\frac{n}{n_1}\right).
   		\end{equation*}
   	\end{lemma}
   	\begin{proof}
   		As a consequence of Theorem~\ref{thm:zetapol} we have
   		\begin{equation*}
   			Z_\C^{(j)}(T)P_\rho^{(j)}\left(T^\frac{n}{n_1}\right)=\frac{P_\C^{(\R,j)}\left(T^{\frac{n}{n_1}}\right)P_\rho^{(j)}\left(T^\frac{n}{n_1}\right)}{\displaystyle \prod_{s=0}^j\left(1-q^{\frac{n}{n_1}s}T^{\frac{n}{n_1}}\right)}=Z_\rho^{(j)}(T)P_\C^{(j)}\left(T^\frac{n}{n_1}\right).
   		\end{equation*}
   		This concludes the proof.
   	\end{proof}
   	
   	The following results generalize~\cite[Theorem~6.5 and Corollaries~6.6,~6.7,~6.8]{byrne2020rank}. The proofs are similar and hence we omit them.
   	
   	\begin{theorem}
   	\label{thm:zetaandpolbeta}
   		For all $\rho\in\{0,\ldots,\frac{n}{n_1}-1\}$ we have
   		\begin{enumerate}[label=(\arabic*)]
   		\setlength\itemsep{3pt}
   			\item $\displaystyle Z_\C^{(\R,j)}(T)=Z_\rho^{(\R,j)}(T)\sum_{a\geq 0}\beta_{\rho,a}^{(\R,j)}(\C)\left(T^\frac{n}{n_1}\right)^a$,
   			\item $\displaystyle P_\C^{(\R,j)}\left(T^{\frac{n}{n_1}}\right)=P_\rho^{(\R,j)}\left(T^{\frac{n}{n_1}}\right)\sum_{a\geq 0}\beta_{\rho,a}^{(\R,j)}(\C)\left(T^\frac{n}{n_1}\right)^a$
   		\end{enumerate}
   		where 
   		\begin{equation*}
   			\beta_{\rho,a}^{(\R,j)}(\C)=\sum_{b=0}^a\frac{p_b^{(\R,j)}(\C)}{\left(p_{\rho,0}^{(\R,j)}\right)^{a-b+1}}\PP_{a-b}\left(p_{\rho,0}^{(\R,j)},-p_{\rho,1}^{(\R,j)},\ldots,-p_{\rho,a-b}^{(\R,j)}\right).
   		\end{equation*}
   	\end{theorem}   
   	
   	\begin{corollary}
   	\label{cor:betaba}
   		For all $a\in\Z_{\geq 0}$ and $\rho\in\{0,\ldots,\frac{n}{n_1}-1\}$, we have
   		\begin{equation*}
   			\beta_{\rho,a}^{(\R,j)}(\C)=\sum_{b=0}^a\frac{b_b^{(\R,j)}(\C)}{\left(b_{\rho,0}^{(\R,j)}\right)^{a-b+1}}\PP_{a-b}\left(b_{\rho,0}^{(\R,j)},-b_{\rho,1}^{(\R,j)},\ldots,-b_{\rho,a-b}^{(\R,j)}\right).
   		\end{equation*}
   	\end{corollary}	
   	
   	\begin{corollary}
   		For all $\rho\in\{0,\ldots,\frac{n}{n_1}-1\}$ we have
   		\begin{equation*}
   			\W_\C^{(\R,j)}(X,Y)=\sum_{a=0}^{n_1-\frac{n_1}{n}t_j^\R}\beta_{\rho,a}^{(\R,j)}\W_{\rho,\frac{n_1}{n}t_j^\R+a}^{(\R,j)}(X,Y).
   	\end{equation*}
   	\end{corollary}
   	\begin{proof}
   		By Theorem~\ref{thm:zetaandpolbeta} and Corollary~\ref{cor:betaba} we have
   		\begin{align*}
   			Z_\C^{(\R,j)}(T)\varphi_{n_1}\left(X^{\frac{n}{n_1}},Y^{\frac{n}{n_1}},T^{\frac{n}{n_1}}\right)&=Z_\rho^{(\R,j)}(T)\varphi_{n_1}\left(X^{\frac{n}{n_1}},Y^{\frac{n}{n_1}},T^{\frac{n}{n_1}}\right)\sum_{b\geq 0}\beta_{\rho,b}^{(\R,j)}(\C)\left(T^\frac{n}{n_1}\right)^b\\
   			&\equiv\sum_{a=0}^{n_1}\W_{\rho,a}^{(\R,j)}(X,Y)T^{n-\frac{n}{n_1}a}\sum_{b\geq 0}\beta_{\rho,b}^{(\R,j)}(\C)T^{\frac{n}{n_1}b}\mod T^{n+1}\\
			&\equiv\sum_{b=0}^{n_1}T^{\frac{n}{n_1}b}\sum_{a=0}^b\beta_{\rho,a}^{(\R,j)}(\C)\W_{\rho,n_1-b+a}^{(\R,j)}(X,Y) \mod T^{n+1}.
   		\end{align*}
   		In particular, the coefficient of $T^{n-t_j^\R}$ is 
   		\begin{equation*}
   			\sum_{a=0}^{n_1-\frac{n_1}{n}t_j^\R}\beta_{\rho,a}^{(\R,j)}\W_{\rho,\frac{n_1}{n}t_j^\R+a}^{(\R,j)}(X,Y)
   		\end{equation*}
   		and the statement follows by Proposition~\ref{prop:varphi}.
   	\end{proof}


\section{Refined Invariants for  Closure-type Anticodes}
\label{sec:refined}
We devote this section to derive a refinement of the tensor binomial moments and weight distribution related to the closure-type anticodes and we establish relations similar to those of Section~\ref{sec:zetaR}. The idea of this refinement is based on the fact that a closure-type anticode is a tensor product of vector spaces. We introduce the concept of \textit{dimension distribution} of a closure-type anticode $A\in\A^\cl$ which we define as the $r$-tuple whose $i$-th component is the dimension of $A^{(i)}$ over $\Fq$. Let $\bdX,\bdY$ and $\bdT$ denote the list of indeterminates $(X_1,\ldots,X_r)$, $(Y_1,\ldots,Y_r)$ and $(T_1,\ldots,T_r)$ respectively.  Throughout this section we consider polynomials that are members of the polynomial rings $\Q[\bdT]$ and $\Q[\bdX,\bdY,\bdT]$. We first introduce some notation.

\begin{notation}
\label{not:prodlat}
	We define the following direct product lattices with respect to the product order.
	\begin{enumerate}[label=(\arabic*)]
	\setlength\itemsep{0.5em}
		\item $\mfZ^\cl:=\Z\times \cdots\times \Z$.
		\item $\mfZ_i^\D:=\begin{cases}
				\{(n_1,\ldots,n_{i-1},a_i,n_{i+1},\ldots,n_r): a_i\in\Z\} & \textup{ if } i\in[r-1],\\[5pt]
				\{(n_1,\ldots,n_{r-1},a_r): a_r\in\Z\} & \textup{ if } i=r \textup{ and }  |\delta_{r}|\neq 1,\\[5pt]
				\emptyset  & \textup{ if } i=r \textup{ and }  |\delta_{r}|= 1.
			\end{cases}$
		\item $\mfZ^\D:=\bigcupdot_{i=1}^{r}\mfZ_i^\D$.
		\item $\mfZ_i^\R:=\begin{cases}
				\{(n_1,\ldots,n_{i-1},a_i,n_{i+1},\ldots,n_r): a_i\in\Z\} & \textup{ if } i\in\delta_{1},\\[5pt]
				\emptyset  & \textup{ if } i\notin\delta_{1}.
			\end{cases}$
		\item $\mfZ^\R:=\bigcupdot_{i=1}^{r}\mfZ_i^\R$.
		\item $\mfN^\cl:=[n_1]\times\cdots\times[n_r]$.
		\item $\mfN_i^\D:=\mfZ_i^\D\cap\mfN^\cl$ and $\mfN^\D:=\mfZ^\D\cap\mfN^\cl$.
		\item $\mfN_i^\R:=\mfZ_i^\R\cap\mfN^\cl$ and $\mfN^\R:=\mfZ^\R\cap\mfN^\cl$.
	\end{enumerate}
	 We denote by $\o$ and $\mathfrak{n}$ the lattice elements $(0,\ldots,0)$ and $(n_1,\ldots,n_r)$ respectively. For any $\a,\b\in\mfZ$ we define $(\a \mod \b)$ to be the element of $\mfZ$ whose $i$-th component is $(\a_i\mod\b_i)$, that is
        \begin{equation*}
        	\a \mod \b=(\a_1 \mod \b_1,\;\ldots,\;\a_r \mod \b_r).
        \end{equation*}
\end{notation}

    
One can easily check that $\mfN^\cl\cup\{\o\}$, $\mfN^\D\cup\{\o\}$ and $\mfN^\R\cup\{\o\}$ with the product order are also lattices. Consider, for example $\mfN^\cl:=[2]\times[3]$. Figures~\ref{fig:lat1} and~\ref{fig:lat2} provide a graph theoretical representation of the lattices $\mfN^\cl$ and $\mfN^\cl\,\cup\{(0,0)\}$ respectively.

	\begin{minipage}{0.5\textwidth}
  			\centering
    		\begin{tikzpicture}
    			\tikzstyle{all nodes}=[inner sep=4pt]
    			\draw node(A) at (0,0){(1,1)}
    					node(B) at (-1,1){(1,2)}
    					node(C) at (1,1){(2,1)}
    					node(D) at (-1,2){(1,3)}
    					node(E) at (1,2){(2,2)}
    					node(F) at (0,3){(2,3)}
    					node(O) at (0,-1){};
    			\draw[-](A)--(B)
    					(A)--(C)
    					(B)--(D)
    					(B)--(E)
    					(C)--(E)
    					(E)--(F)
    					(D)--(F); 
    		\end{tikzpicture}
    		\captionsetup{font=footnotesize}
			\captionof{figure}{\label{fig:lat1} Lattice $\mfN^\cl$.}
    	\end{minipage}
		\begin{minipage}{0.5\textwidth}
  			\centering
    		\begin{tikzpicture}
    			\tikzstyle{all nodes}=[inner sep=4pt]
    			\draw node(A) at (0,0){(1,1)}
    					node(B) at (-1,1){(1,2)}
    					node(C) at (1,1){(2,1)}
    					node(D) at (-1,2){(1,3)}
    					node(E) at (1,2){(2,2)}
    					node(F) at (0,3){(2,3)}
    					node(O) at (0,-1){(0,0)};
    			\draw[-](A)--(B)
    					(A)--(C)
    					(B)--(D)
    					(B)--(E)
    					(C)--(E)
    					(E)--(F)
    					(D)--(F)
    					(O)--(A); 
    		\end{tikzpicture}
    		\captionsetup{font=footnotesize}
			\captionof{figure}{\label{fig:lat2} Lattice $\mfN^\cl\cup\{(0,0)\}$.}
    	\end{minipage}
    	
\begin{definition}
	Let $A\in\A^\cl$ be a non-zero closure-type anticode. We define the \textbf{dimension distribution} $\dd(A)$ of $A$ to be  
	\begin{equation*}
        \dd(A):=\left(\dimq{A^{(1)}},\cdots,\dimq{A^{(r)}}\right)\in\mfN^\cl
    \end{equation*}
    We set $\dd(\<0\>_{\Fq})=\o$.
\end{definition}

    \begin{remark}
    	We have the following.
       \begin{enumerate}[label=(\arabic*)]
       \setlength\itemsep{0.5em}
       \item $A\in\A^\cl$ if and only if $\dd(A)\in\mfN^\cl\,\cup\{\o\}$.
       \item $A\in\A^\D$ if and only if $\dd(A)\in\mfN^\D\,\cup\{\o\}$.
       \item $A\in\A^\R$ if and only if $\dd(A)\in\mfN^\R\,\cup\{\o\}$.
       \end{enumerate}
    \end{remark}
    
    \begin{notation}
      For any $\a\in\mfN\cup\{\o\}$, we let $\A^\cl_{\a}:=\left\{A\in\A^\cl:\dd(A)=\a\right\}$ and $\[\a\]$ to be the product of the components of $\a$, that is $\[\a\]:=\a_1\cdots\a_r$.
    \end{notation}

We now give a refined version of the invariants related to closure-type anticodes based on the structure of the direct product lattice of this set of anticodes.

    \begin{definition}
        For any $\a\in\mfN^\cl\cup\{\o\}$, the $(\a,j)$\textbf{-th refined  tensor binomial moment} related to $\A_\a^\cl$ of $\C$ is
        \begin{equation*}
            B_\a^{(\cl,j)}(\C):=\sum_{A\in\A_\a^\cl}B_A^{(\cl,j)}(\C),\qquad\textup{ where }\qquad B_A^{(\cl,j)}(\C):=\qbin{\dimq{\C\cap A}}{j}
        \end{equation*}
        for any $A\in\A_\a^\cl$.
    \end{definition}
	
	\begin{definition}
        For any $\a\in\mfN^\cl\cup\{\o\}$, we define        
        \begin{equation*}
            W_\a^{(\cl, j)}(\C):=\sum_{A\in\A_\a^\cl}W_A^{(\cl,j)}(\C),
        \end{equation*}
        where $W_A^{(\cl,j)}:=|\{\mD\leq(\C\cap A):\dimq{\mD}=j \textup{ and } \mD\leq B\leq A,B\in\A^\cl\Longrightarrow B=A\}|$, for any $A\in\A_\a^\cl$. We refer to the set $\{ W_\a^{(\cl, j)}(\C):\a\in\mfN^\cl\cup\{\o\}\}$ as the $j$-\textbf{th refined tensor weight distribution} related to $\A^\cl$ of $\C$.
    \end{definition}
    
    Clearly, for all $a\in[n]$ we have 
    \begin{equation*}
    		B_a^{(\cl,j)}(\C)=\sum_{\substack{\a\in\mfN^\cl\\[2pt]\[\a\]=a}} B_\a^{(\cl,j)}(\C) \qquad\textup{ and }\qquad W_a^{(\cl,j)}(\C)=\sum_{\substack{\a\in\mfN^\cl\\[2pt]\[\a\]=a}} W_\a^{(\cl,j)}(\C) 
    	\end{equation*}

The following extends Proposition~\ref{prop:oldBaWb}.
    
    \begin{theorem}
    \label{thm:BaWbcl}
        The following hold for any $\a,\b\in\mfN^\cl\cup\{\o\}$ .
        \begin{enumerate}[label=(\arabic*)]
  			\item $\displaystyle B_\a^{(\cl,j)}(\C)=\sum_{\substack{\b\in\mfN^\cl\\[1pt]\b\leq\a}}W_\b^{(j)}(\C)\prod_{i=1}^r\qbin{n_i-\b_i}{\a_i-\b_i}$.
            \item $\displaystyle W_\b^{(\cl,j)}(\C)=\sum_{\substack{\a\in\mfN^\cl\\[1pt]\a\leq\b}}B_\a^{(\cl,j)}(\C)\prod_{i=1}^r(-1)^{\b_i-\a_i}q^{\binom{\b_i-\a_i}{2}}\qbin{n_i-\a_i}{\b_i-\a_i}$.
  		\end{enumerate}
    \end{theorem}
    \begin{proof}
        For ease of notation, in the remainder of this proof we write $B_\a^{(j)}$ and $W_\b^{(j)}$ instead of $B_\a^{(\cl,j)}(\C)$ and $W_\b^{(\cl,j)}(\C)$ respectively. Observe that $B_\o^{(j)}(\C)=W_\o^{(j)}(\C)=0$ by definition, since $j$ is at least $1$. Therefore, we can assume $\a,\b\in\mfN$ for the remainder of the proof. Observe that for any $A\in\A_\a^\cl$ we have 
        \begin{equation}
        \label{eq1:mob}
        	\sum_{\substack{A'\in\A_\b^\cl\\[1pt]A'\leq A}} W_{A'}^{(j)}=|\{\mD\leq\C:\dimq{\mD}=j\textup{ and }\mD\leq A\}|= B_A^{(j)},
        \end{equation}
        and by the M\"obius inversion formula we obtain
        \begin{equation}
        \label{eq2:mob}
        	W_{A'}^{(j)}=\sum_{\tiny\begin{matrix}A\in\A^\cl\\A\leq A'\end{matrix}}\prod_{i=1}^r(-1)^{\dimq{A'^{(i)}}-\dimq{A^{(i)}}}q^{\binom{\dimq{A'^{(i)}}-\dimq{A^{(i)}}}{2}}B_A^{(j)},
        \end{equation}
        since $\A_\a^\cl$ is a product lattice. Therefore, by Equation~\eqref{eq1:mob}, we get
       \begin{align*}
            B_\a^{(j)}&=\sum_{A\in\A_\a^\cl}B_A^{(j)}=\sum_{A\in\A_\a^\cl}\left(\underbrace{W_{\<0\>_{\Fq}}^{(j)}}_{=\,0}+\sum_{\substack{A'\in\A^\cl\\[1pt]A'\leq A}} W_{A'}^{(j)}\right)\\
            &=\sum_{\substack{\b\in\mfN^\cl\\[1pt]\b\leq_\mfN\a}}\sum_{A'\in\A_\b^\cl}W_{A'}^{(j)}|\{A\in\A_\a^\cl:A'\leq A\}|\\
            &=\sum_{\substack{\b\in\mfN^\cl\\[1pt]\b\leq_\mfN\a}}\prod_{i=1}^r\qbin{n_i-\b_i}{\a_i-\b_i}\sum_{A'\in\A_\b^\cl}W_{A'}^{(j)}\\
            &=\sum_{\substack{\b\in\mfN^\cl\\[1pt]\b\leq_\mfN\a}}\prod_{i=1}^r\qbin{n_i-\b_i}{\a_i-\b_i}W_{\b}^{(j)},
        \end{align*}
        where the last equality follows by~\cite[Lemma~4.9]{byrne2021tensor}. In order to prove the second part of statement, observe that by Equation~\eqref{eq2:mob} we have
        \begin{align*}
            W_\b^{(j)}&=\sum_{A'\in\A_\b^\cl}W_{A'}^{(j)}\\
            &=\sum_{A'\in\A_\b^\cl}\left(\underbrace{B_{\<0\>_{\Fq}}^{(j)}\prod_{i=1}^r(-1)^{\b_i}q^{\binom{\b_i}{2}}}_{=\,0}+\sum_{\tiny\begin{matrix}A\in\A^\cl\\A\leq A'\end{matrix}}B_A^{(j)}\prod_{i=1}^r(-1)^{\b_i-\dimq{A^{(i)}}}q^{\binom{\b_i-\dimq{A^{(i)}}}{2}}\right)\\
            &=\sum_{\substack{\a\in\mfN^\cl\\[1pt]\a\leq_\mfN\b}}\prod_{i=1}^r(-1)^{\b_i-\a_i}q^{\binom{\b_i-\a_i}{2}}\sum_{A\in\A_\a^\cl}B_\a^{(j)}\left|\left\{A\in\A_\b,A\leq A'\right\}\right|\\
            &=\sum_{\substack{\a\in\mfN^\cl\\[1pt]\a\leq_\mfN\b}}B_\a^{(j)}(\C)\prod_{i=1}^r(-1)^{\b_i-\a_i}q^{\binom{\b_i-\a_i}{2}}\prod_{i=1}^r\qbin{n_i-\a_i}{\b_i-\a_i}
        \end{align*}
        where the last equality follows again by Lemma~\cite[Lemma~4.9]{byrne2021tensor}.
    \end{proof} 
    
Clearly, the results derived so far in this section apply also for the invariants related to $\A^\D$ and $\A^\R$ since the latter are proper subclasses of $\A^\cl$ as observed in~\cite{byrne2021tensor}. Observe that, for any $\a\in\mfN^\cl\cup\{\o\}$, we have $B_\a^\D(\C)=W_\a^\D(\C)=0$ if $\a\notin\mfN^\D$ and $B_\a^\R(\C)=W_\a^\R(\C)=0$ if $\a\notin\mfN^\R$. 

\begin{notation}
	For the rest of this section, we let $\P$ denotes the type of the family of  anticodes considered, in particular, $\P\in\{\cl,\D,\R\}$.
\end{notation}

 The following result is the analogue of Proposition~\ref{prop:Badetermied}.
    
    \begin{theorem}
        The following holds for all $\a\in\mfN^\P\cup\{\o\}$.
        \begin{equation*}
            B_\a^{(\P,j)}(\C)=
            \begin{cases}
                 0 & \displaystyle\textup{ if } \[\a\]<t_j^\P,\\[2ex]
                 \displaystyle\qbin{ k+\[\a\]-n}{j}\prod_{i=1}^r\qbin{n_i}{\a_i} & \textup{ if }   \displaystyle \[\a\]>n-(s_1^\P)^\perp.
            \end{cases}
        \end{equation*}
    \end{theorem}   
    \begin{proof}
    	An argument similar to the one in the proof of~\cite[Theorem~6.6]{byrne2021tensor} shows that
    	\begin{equation*}
    		B_\a^{(\P,j)}(\C)=
            \begin{cases}
                 0 & \displaystyle\textup{ if } \[\a\]<t_j^\P,\\[2ex]
                 \displaystyle\qbin{ k+\[\a\]-n}{j}\left|\A_{\a}^\P\right| & \textup{ if }   \displaystyle \[\a\]>n-(s_1^\P)^\perp.
            \end{cases}
    	\end{equation*}
    	and the statement follows from the fact that $\A_\a^\P$ is a product lattice.
    \end{proof}
     	
In what follows, we introduce a refinement of the normalized binomial moments and a multivariate version of the tensor weight enumerator and tensor zeta function with respect to $\A^\P$. We stress the fact that such definitions strongly depends on the family of anticodes considered.
    
    \begin{definition}
    \label{def:norbcl}
       For any $\a\in\mathfrak{Z}^\P$, the $(\a,j)$-\textbf{th refined normalized tensor binomial moment} of $\C$ with respect to $\A^\P$ is
        \begin{equation*}
            b_{\a}^{(\P,j)}:=
            \begin{cases}
                0 & \displaystyle\textup{ if } \[\a\]<t_j^\P,\\[2ex]
                 \displaystyle\frac{B_\a^{(\P,j)(\C)}}{\displaystyle\prod_{i=1}^r\qbin{n_i}{\a_i}}  & \textup{ if }  \displaystyle t_j^\P\leq\[\a\]\leq n-(s_1^\P)^\perp \textup{ and } \mQ_\P,\\[8ex]
                 \displaystyle\qbin{k+\[\a\]-n}{j} & \textup{ if }   \displaystyle \[\a\]>n-(s_1^\P)^\perp \textup{ and } \mQ_\P,\\[5ex]
                 0 & \textup{ if }   \displaystyle \[\a\]>n-(s_1^\P)^\perp \textup{ and } \neg\mQ_\P,
            \end{cases}
        \end{equation*}
        where $\mQ_\P$ is the predicate defined as follows
        \begin{equation*}
        	\mQ_\P:=\begin{cases}
        		\a\in\mfN^\cl & \textup{ if } \a\leq\n\textup{ and }\P=\cl,\\[2pt]
        		\n<\a & \textup{ if } \n<\a \textup{ and }\P=\cl,\\[2pt]
        		n\mid \[\a\]\, m \textup{ for some } m\in\Delta & \textup{ if } \P=\D,\\[2pt]
			n\mid \[\a\]\,n_1 & \textup{ if } \P=\R.
        	\end{cases}
        \end{equation*}
    \end{definition}

Note that that in this case we did not shift the indices as in Definition~\ref{def:ba}. Moreover, if $t_j^\P\leq\[\a\]\leq n-(s_1^\P)^\perp$ and $\a\notin\mfN^\P$ then $b_{\a}^{(\P,j)}=0$. 

     \begin{definition}
        The $j$\textbf{-th refined tensor weight enumerator} related to $\A^\P$ is
        \begin{align*}
            \W_\C^{(\P,j)}(\bdX,\bdY):=\sum_{\a\in\mfN^\P}W_\a^{(\P,j)}(\C)\prod_{i=1}^rX_i^{n_i-\a_i}Y_i^{\a_i}
            =\sum_{\substack{\a\in\mfN^\P\\[1pt]\[\a\]\geq t_j^\P}}W_\a^{(\P,j)}(\C)\prod_{i=1}^rX_i^{n_i-\a_i}Y_i^{\a_i}.
        \end{align*}
    \end{definition}
    
    It is not hard to check that the set $\left\{\prod_{i=1}^rB_{n_i,\a_i}(X_i,Y_i;q):\a\in\mfN^\P\right\}$ of products of $q$-Bernstein polynomials is a $\Q$-basis of the space of the refined generalized tensor weight enumerators related to $\A^\P$. The following result computes the coefficients of the multivariate polynomial $\W_\C^{(\P,j)}(\bdX,\bdY)$ with respect such a basis.
    
    \begin{proposition}
    \label{prop:WclB}
        The following holds.
        \begin{equation*}
            \W_\C^{(\P,j)}(\bdX,\bdY):=\sum_{\substack{\a\in\mfN^\P\\ \[\a\]\geq t_j^\P}}b_{\a}^{(\P,j)}(\C)\prod_{i=1}^rB_{n_i,\a_i}(X_i,Y_i;q).
        \end{equation*}
    \end{proposition}
    \begin{proof}
        Applying the inversion formula of the $q$-Bernstein polynomial to the definition of the refined generalized tensor weight enumerator, we get
        \begin{align*}
           \W_\C^{(\P,j)}(\bdX,\bdY)&=\sum_{\substack{\b\in\mfN^\P\\ \[\b\]\geq t_j^\P}}W_\b^{(\P,j)}(\C)\prod_{i=1}^rX_i^{n_i-\b_i}Y_i^{\b_i}\\
            &=\sum_{\substack{\b\in\mfN^\P\\ \[\b\]\geq t_j^\P}}W_\b^{(\P,j)}(\C)\prod_{i=1}^r\qbin{n_i}{\b_i}^{-1}\sum_{\a_i=\b_i}^{n_i}\qbin{\a_i}{\b_i}B_{n_i,\a_i}(X_i,Y_i;q)\\
            &=\sum_{\substack{\b\in\mfN^\P\\ \[\b\]\geq t_j^\P}}W_\b^{(\P,j)}(\C)\left(\prod_{i=1}^r\qbin{n_i}{\b_i}^{-1}\right)\sum_{\substack{\a\in\mfN^\P\\\b\leq\a}}\prod_{i=1}^r\qbin{\a_i}{\b_i}B_{n_i,\a_i}(X_i,Y_i;q)\\
            &=\sum_{\substack{\a\in\mfN^\P\\ \[\a\]\geq t_j^\P}}\left(\prod_{i=1}^rB_{n_i,\a_i}(X_i,Y_i;q)\right)\sum_{\substack{\a\in\mfN^\P\\\b\leq\a\\ \[\b\]\geq t_j^\P}}W_\b^{(\P,j)}(\C)\prod_{i=1}^r\qbin{n_i}{\b_i}^{-1}\qbin{\a_i}{\b_i}\\
            &=\sum_{\substack{\a\in\mfN^\P\\ \[\a\]\geq t_j^\P}}\left(\prod_{i=1}^rB_{n_i,\a_i}(X_i,Y_i;q)\qbin{n_i}{\a_i}^{-1}\right)\sum_{\substack{\a\in\mfN^\P\\\b\leq\a\\ \[\b\]\geq t_j^\P}}W_\b^{(\cl,j)}(\C)\prod_{i=1}^r\qbin{n_i-\b_i}{\a_i-\b_i}\\
            &=\sum_{\substack{\a\in\mfN^\P\\ \[\a\]\geq t_j^\P}}b_{\a}^{(\cl,j)}(\C)\prod_{i=1}^rB_{n_i,\a_i}(X_i,Y_i;q)
        \end{align*}
    \end{proof} 
    
    \begin{definition}
	The $j$\textbf{-th refined tensor zeta function} related to $\A^\P$ of $\C$ is 
       \begin{equation*}
           Z_{\C}^{(\P,j)}(\bdT):=\sum_{\substack{\a\in\mathfrak{Z}^\P\\\mQ_\P}}b_\a^{(\P,j)}(\C)\prod_{i=1}^rT_i^{\a_i}=\sum_{\substack{\a\in\mathfrak{Z}^\P\\{\[\a\]}\geq t_j^\P\\\mQ_\P}} b_\a^{(\P,j)}(\C)\prod_{i=1}^rT_i^{\a_i}.
       \end{equation*}
	\end{definition}

	The following extends Proposition~\ref{prop:varphi}.
	
	\begin{proposition}
	\label{prop:varphicl}
		Define the function
		\begin{equation*}
			\varphi_{\a}^\P(\bdX,\bdY,\bdT):=\sum_{\substack{\b\in\mfN^\P\\\b\leq\a}}\prod_{i=1}^rB_{a_i,\b_i}(X_i,Y_i;q)T_i^{a_i-\b_i}.
		\end{equation*}
		We have that $\W_\C^{(\P,j)}(\bdX,\bdY)$ is the coefficient of $\prod_{i=1}^rT_i^{n_i}$ in the expression $Z_{\C}^{(\P,j)}(\bdT)\varphi_{\n}^\P(\bdX,\bdY,\bdT)$.
	\end{proposition}
	\begin{proof}
		The following hold.
		\begin{align*}
			&Z_{\C}^{(\P,j)}(\bdT)\varphi_{\n}^\P(\bdX,\bdY,\bdT)\\&=\left(\sum_{\substack{\a\in\mathfrak{Z}^\P\\{\[\a\]}\geq t_j^\cl\\\mQ_\P}} b_\a^{(\P,j)}(\C)\prod_{i=1}^rT_i^{\a_i}\right)\left(\sum_{\b\in\mfN^\P}\prod_{i=1}^rB_{n_i,\b_i}(X_i,Y_i;q)T_i^{n_i-\b_i}\right)\\
			&=\sum_{\substack{\a\in\mathfrak{Z}^\P\\{\[\a\]}\geq t_j^\P\\\mQ_\P}}\sum_{\b\in\mfN^\P}b_\a^{(\P,j)}(\C)\prod_{i=1}^rB_{n_i,\b_i}(X_i,Y_i;q)\prod_{i=1}^rT_i^{n_i-\b_i+\a_i}\\
			&\equiv\sum_{\substack{\b\in\mfN^\P\\{\[\b\]}\geq t_j^\P}}\prod_{i=1}^rT_i^{\b_i}\sum_{\substack{\a\in\mfN^\P\\\a\leq\b\\ \[\a\]\geq t_j^\P}}b_\a^{(\cl,j)}\prod_{i=1}^rB_{n_i,n_i-\b_i+\a_i}(X_i,Y_i;q) \mod \<T_1^{n_1+1},\ldots,T_r^{n_r+1}\>_\Q .
		\end{align*}
		The result follows by Proposition~\ref{prop:WclB}.
	\end{proof}

\begin{example}
	Let $\C$ be the code in Example~\ref{ex:1-TMRD} and one can check that its generalized tensor weights with respect to $\A^\cl$ are 
	\begin{equation*}
	\begin{array}{lll}
		t_1^\cl(\C)=4, &\qquad t_2^\cl(\C)=9, &\qquad   t_3^\cl(\C)=12,\\[8pt]
		t_4^\cl(\C)=t_5^\cl(\C)=16, &\qquad t_6^\cl(\C)=t_7^\cl(\C)=18, &\qquad t_8^\cl(\C)=\cdots=t_13^\cl(\C)=24.
	\end{array}
	\end{equation*}
	Moreover, the following hold.
	\begin{align*}
		\varphi_{(2,3,4)}^\cl(\bdX,\bdY,\bdT)=&\,Y_1^2Y_2^3Y_3^4+(3X_1Y_1Y_2^3Y_3^4-3Y_1^2Y_2^3Y_3^4)T_1\\
		&+(7X_2Y_1^2Y_2^2Y_3^4-7Y_1^2Y_2^3Y_3^4)T_2+(15X_3Y_1^2Y_2^3Y_3^3-15Y_1^2Y_2^3Y_3^4)T_3\\
		&+(21X_1X_2Y_1Y_2^2Y_3^4-21X_1Y_1Y_2^3Y_3^4-21X_2Y_1^2Y_2^2Y_3^4 + 21Y_1^2Y_2^3Y_3^4)T_1T_2\\
		&+(7X_2^2Y_1^2Y_2Y_3^4 - 21X_2Y_1^2Y_2^2Y_3^4 + 14Y_1^2Y_2^3Y_3^4)T_2^2+\cdots\\[8pt]
		Z_\C^{(\cl,1)}(\bdT)=&\,\frac{4}{735}T_1T_2^2T_3^2+\frac{4}{105}T_1T_2^2T_3^3+\frac{4}{21}T_1T_2^2T_3^4+\frac{4}{105}T_1T_2^3T_3^2+\frac{17}{45}T_1T_2^3T_3^3\\
		&+3T_1T_2^3T_3^4+\frac{1}{245}T_1^2T_2T_3^2+\frac{1}{35}T_1^2T_2T_3^3+\frac{1}{7}T_1^2T_2T_3^4+\cdots
		\\[8pt]
		\W_\C^{(\cl,1)}(\bdX,\bdY)=&\,6374\,Y_1^2Y_2^3Y_3^4+1556\,X_3Y_1^2Y_2^3Y_3^3+41\,X_3^2Y_1^2Y_2^3Y_3^2+62\,X_2Y_1^2Y_2^2Y_3^4\\
		&+121\,X_2X_3Y_1^2Y_2^2Y_3^3+26\,X_2X_3^2Y_1^2Y_2^2Y_3^2+X_2X_3^3Y_1^2Y_2^2Y_3\\
			&+X_2^2X_3^2Y_1^2Y_2Y_3^2+5X_1X_3Y_1Y_2^3Y_3^3+4X_1X_2X_3^2Y_1Y_2^2Y_3^2.
	\end{align*}
	Therefore, $\W_\C^{(\cl,1)}(\bdX,\bdY)$ is the coefficient of $T_1^2T_2^3T_3^4$ in the expression $Z_\C^{(\cl,1)}(\bdT)\varphi_{(2,3,4)}^\cl(\bdX,\bdY,\bdT)$ as predicted by Proposition~\ref{prop:varphicl}.
\end{example}

We now establishing the MacWilliams-type identity for the refined binomial moments related to $\A^\D$ and $\A^\R$. We need the following preparatory lemma which can be seen as the analogue of~\cite[Theorem~6.7]{byrne2021tensor}. We omit the proof as it is similar to the one of ~\cite[Theorem~6.7]{byrne2021tensor}.

  	\begin{lemma}
  		\label{lem:BaBadual}
  		The following hold for any $\a\in\mfN^\P\cup\{\mathfrak{o}\}$.
  		\begin{equation*}
  			B_\a^{(\P,j)}(\C)=\sum_{s=0}^jq^{s(k+\[\a\]-n-j+s)}\qbin{k+\[\a\]-n}{j-s}\sum_{A\in\A_\a^\P}\qbin{\dimq{\C^\perp\cap A^\perp}}{s}.
  		\end{equation*}
  	\end{lemma}
  	
\begin{theorem}
	The following hold.
	\begin{enumerate}[label=(\arabic*)]
		\item\label{item1:macwill} $\displaystyle B_\a^{(\D,j)}(\C)=\sum_{s=0}^jq^{s(k+\[\a\]-n-j+s)}\qbin{k+\[\a\]-n}{j-s}B_{\overline{\a}}^{(\D,s)}(\C^\perp)$ for all $\mfN^\D\cup\{\o\}$,
		\item\label{item2:macwill} $\displaystyle B_\a^{(\R,j)}(\C)=\sum_{s=0}^jq^{s(k+\[\a\]-n-j+s)}\qbin{k+\[\a\]-n}{j-s}B_{\overline{\a}}^{(\R,s)}(\C^\perp)$ for all $\mfN^\R\cup\{\o\}$
	\end{enumerate}
	where $\overline{\a}=\o$ if $\a=\n$ and $\overline{\a}=\n-(\a\mod\n)$ otherwise.
\end{theorem}
\begin{proof}
	We only prove~\eqref{item1:macwill} as the proof of~\eqref{item2:macwill} is similar. By Lemma~\ref{lem:BaBadual}, we have
		\begin{equation*}
  			B_\a^{(D,j)}(\C)=\sum_{s=0}^jq^{s(k+\[\a\]-n-j+s)}\qbin{k+\[\a\]-n}{j-s}\sum_{A\in\A_\a}\qbin{\dimq{\C^\perp\cap A^\perp}}{s}
  		\end{equation*}
  		for any $\a\in\mfN^\D\cup\{\o\}$, since $\A^\D\subseteq\A^\cl$ and $\mfN^\D\subseteq\mfN^\cl$. Recall that $\A^\D$ is closed under duality, by Proposition~\ref{prop:charoverlineA}, and it is not hard to see that for any $A\in\A_\a^\D$ we have $\dd(A^\perp)=\overline{\a}$ where $\overline{\a}$ is defined as in the statement. Hence, the map $:\A_{\a}^\D\longrightarrow\A_{\overline{\a}}^\D:A\longmapsto A^\perp$ is an isomorphism. This implies
  		\begin{equation*}
  			\sum_{A\in\A_\a}\qbin{\dimq{\C^\perp\cap A^\perp}}{s}=\sum_{A\in\A_{\overline{\a}}}\qbin{\dimq{\C^\perp\cap A}}{s}=B_{\overline{\a}}^{(\D,s)}.
  		\end{equation*}
  		for all $s\in[j]\cup\{0\}$. The statement follows.
\end{proof}

We conclude this section providing the analogue of Theorem~\ref{thm:zetapol} for the refined zeta functions associated to Ravagnani-type anticodes. 

\begin{theorem}
\label{thm:zetapolRrefined}
	For any $i\in\{1,\ldots,|\delta_1|\}$, there exist unique polynomial $P_{\C,i}^{(\R,j)}$ such that $Z_\C^{(\R,j)}(\bdT)$ is equal to
	\begin{equation*}
		\left(\prod_{\ell=1}^{|\delta_1|}\prod_{s=0}^j\left(1-q^{s\frac{n}{n_1}}T_\ell\right)\right)^{-1}\sum_{i=1}^{\delta_1}P_{\C,i}^{(\R,j)}\left(\prod_{\substack{\ell=1\\\ell\neq i}}^{|\delta_1|}\prod_{s=0}^j\left(1-q^{s\frac{n}{n_1}}T_\ell\right)\right)-(|\delta_1|-1)b_{\n}^{(\R,j)}(\C)\prod_{\ell=1}^{r}T_\ell^{n_\ell}.
	\end{equation*}
	In particular, we have 
	\begin{equation*}
		P_{\C,i}^{(\R,j)}(\bdT)=\left(\sum_{a=0}^bp_{\a_i(a)}^{(\R,j)}(\C)T_i^a\right)\prod_{\substack{\ell=1\\\ell\neq i}}^rT_\ell^{n_\ell}
	\end{equation*}
	where $b:=n_1-\frac{n_1}{n}t_j^\R+j+1$, $\a_i(a)$ is the element of $\mfZ^\R$ with $a$ in position $i$ and 
	\begin{equation}
	\label{eq:parefined}
		p_{\a_i(a)}^{(\R,j)}(\C)=\sum_{s=0}^{j+1}\bbin{j+1}{s}{\frac{n}{n_1}}(-1)^sq^{\frac{n}{n_1}\binom{s}{2}}b_{\a_i(a-s)}^{(\R,j)}(\C).
	\end{equation}
\end{theorem}
\begin{proof}
	Define, for any  $i\in\{1,\ldots,|\delta_1|\}$, the formal power series
	\begin{equation*}
		Z_{\C,i}^{(\R,j)}(\bdT):=\left(\sum_{a\geq 0}b_{\a_i(a)}^{(\R,j)}(\C)T_i^a\right)\prod_{\substack{\ell=1\\\ell\neq i}}^rT_\ell^{n_\ell}.
	\end{equation*}
	One can check that by definition of Ravagnani-type anticodes and the underlying lattice of dimension distribution we have
	\begin{equation}
	\label{eq:zetarefined}
		Z_\C^{(\R,j)}(\bdT)=\sum_{i\in\{1,\ldots,|\delta_1|\}}Z_{\C,i}^{(\R,j)}(\bdT)-(|\delta_1|-1)b_{\n}^{(\R,j)}(\C)\prod_{\ell=1}^{r}T_\ell^{n_\ell}.
	\end{equation}
	The remainder of the proof is similar to the one of Theorem~\ref{thm:zetapol}. In particular, applying Lemma~\ref{lem:bin} to the power series
	\begin{equation}
	\label{eq:zetapolrefined}
		P_{\C,i}^{(\R,j)}(\bdT):=Z_\C^{(\R,j)}(\bdT)\prod_{s=0}^j\left(1-q^{s\frac{n}{n_1}}T_i\right)
	\end{equation}
	gives the coefficients $p_{\a_i(a)}^{(\R,j)}(\C)$ as in~\eqref{eq:parefined}. Moreover, a standard computation using Lemma~\ref{lem:bin} shows that
	\begin{equation*}
		\sum_{s=0}^{j+1}\bbin{j+1}{s}{\frac{n}{n_1}}(-1)^sq^{\frac{n}{n_1}\binom{s}{2}}b_{\a_i(a-s)}^{(\R,j)}(\C)=0
	\end{equation*}
	for all $a\notin\{0,\ldots,n_1-\frac{n_1}{n}t_j^\R+j+1\}$. The statement follows by combining~\eqref{eq:zetarefined} and~\eqref{eq:zetapolrefined}.
\end{proof}

\begin{example}
	Let $\C$ be the code as in Example~\ref{ex:1-TMRDm|k}. We have
	\begin{align*}
		Z_\C^{(\R,1)}=&\, 4095 T_1^3T_2^3T_3^4+16777215(T_1^4T_2^3T_3^4+T_1^3T_2^4T_3^4)\\
		&\, +68719476735(T_1^5T_2^3T_3^4+T_1^3T_2^5T_3^4)+\cdots\\[8pt]
		Z_{\C,1}^{(\R,1)}=&\, 4095 T_1^3T_2^3T_3^4+16777215T_1^4T_2^3T_3^4+68719476735T_1^5T_2^3T_3^4+\cdots\\[8pt]
		Z_{\C,2}^{(\R,1)}=&\, 4095 T_1^3T_2^3T_3^4+16777215T_1^3T_2^4T_3^4+68719476735T_1^3T_2^5T_3^4+\cdots
	\end{align*}
	and therefore 
	\begin{equation*}
		Z_\C^{(\R,1)}=Z_{\C_1}^{(\R,1)}+Z_{\C_2}^{(\R,1)}-4095 T_1^3T_2^3T_3^4.
	\end{equation*}
	Moreover, define
	\begin{equation*}
		P_{\C,1}^{(\R,1)}:=4095 T_1^3T_2^3T_3^4 \qquad \textup{ and }\qquad P_{\C,2}^{(\R,1)}:=4095 T_1^3T_2^3T_3^4,
	\end{equation*} 
	and one can check that
	\begin{equation*}
		Z_\C^{(\R,1)}=\frac{P_{\C,1}^{(\R,1)}}{(1-T_1)(1-q^{12}T_1)}+\frac{P_{\C,3}^{(\R,1)}}{(1-T_1)(1-q^{12}T_1)}-4095 T_1^3T_2^3T_3^4
	\end{equation*}
	as predicted by Theorem~\ref{thm:zetapolRrefined}.
\end{example}


\section{Connections}
\label{sec:bounds}
In this section we show an application of theory developed in the previous sections. We derive new connections between the tensor weights associated to $\A^\D$ and $\A^\R$ which establish relations between the latter and the refined tensor zeta functions. An inequality involving the $j$-th tensor weight related to $\A^\R$ is derived in the next result. As immediate consequence of this result we have a generalization of~\cite[Theorem~4]{blanco2018rank}. We provide a more concise proof using the notion of normalized tensor binomial moments related to $\A^\R$ (see Definition~\ref{def:ba}).

\begin{theorem}
\label{thm:boundR}
	If $t_j^\R< n$ then 
	\begin{equation*}
		t_j^\R\leq \frac{n}{n_1}\log_q\left(\frac{b_{\frac{n}{n_1}}^{(j)}}{b_0^{(j)}}(q-1)+1\right)-\frac{n}{n_1}.
	\end{equation*}
\end{theorem} 
\begin{proof}
	For ease of notation, throughout this proof we write $W_b^{(j)}$ and $B_b^{(j)}$ instead of $W_b^{(\R,j)}(\C)$ and $B_b^{(\R,j)}(\C)$, for all $b\in[n]\cap\{0\}$, and $b_a^{(j)}$ and $t_j$ instead of $b_a^{(\R,j)}(\C)$ and $t_j^\R$, for all $a\in\Z$. Proposition~\ref{prop:BaWb} and the fact that $B_a^{(j)}=0$, for all $a\in\{0,\ldots,t_j^\R-1\}$ and $a\in\{t_j^\R,\ldots,n\}$ such that $n\nmid an_i$, imply
	\begin{align*}
		W_{t_j^\R+\frac{n}{n_1}}^{(j)}&=B_{t_j^\R+\frac{n}{n_1}}^{(j)}|\delta_{1}|-B_{t_j^\R}^{(j)}|\delta_{1}|\qbin{n_1-t_j\frac{n_1}{n}}{1}\\
		&=b_{\frac{n}{n_1}}^{(j)}|\delta_{1}|^2\qbin{n_1}{t_j\frac{n_1}{n}+1}-b_0^{(j)}|\delta_{1}|^2\qbin{n_1-t_j\frac{n_1}{n}}{1}\qbin{n_1}{t_j\frac{n_1}{n}}
	\end{align*}
	where the last equality follows from Definition~\ref{def:ba}. Observe that $W_{t_j^\R+\frac{n}{n_1}}^{(j)}\geq 0$ and $|\delta_{1}|\geq 0$ and we have
	\begin{align*}
		0&\leq  b_{\frac{n}{n_1}}^{(j)}\qbin{n_1}{t_j\frac{n_1}{n}+1}-b_0^{(j)}\qbin{n_1-t_j^\R\frac{n_1}{n}}{1}\qbin{n_1}{t_j\frac{n_1}{n}}\\
		&=b_{\frac{n}{n_1}}^{(j)}\frac{q^{n_1-t_j\frac{n_1}{n}}-1}{q^{t_j\frac{n_1}{n}+1}-1}\qbin{n_1}{t_j\frac{n_1}{n}}-b_0^{(j)}\qbin{n_1-t_j\frac{n_1}{n}}{1}\qbin{n_1}{t_j\frac{n_1}{n}}.
	\end{align*}
	Therefore, since  $\qbin{n_1}{t_j\frac{n_1}{n}}> 0$ and $b_0^{(j)}> 0$, we get
	\begin{equation}
	\label{eq1:boundR}
		0\leq \frac{b_{\frac{n}{n_1}}^{(j)}}{b_0^{(j)}}\cdot\frac{q^{n_1-t_j\frac{n_1}{n}}-1}{q^{t_j^\R\frac{n_1}{n}+1}-1}-\qbin{n_1-t_j\frac{n_1}{n}}{1}=\frac{b_{\frac{n}{n_1}}^{(j)}}{b_0^{(j)}}\cdot\frac{q^{n_1-t_j\frac{n_1}{n}}-1}{q^{t_j\frac{n_1}{n}+1}-1}-\frac{q^{n_1-t_j\frac{n_1}{n}}-1}{q-1}
	\end{equation}
	which implies
	\begin{equation*}
		q^{t_j\frac{n_1}{n}+1}\leq \frac{b_{\frac{n}{n_1}}^{(j)}}{b_0^{(j)}}(q-1)+1.
	\end{equation*}
	The statement follows.
\end{proof}

The following is a generalization of~\cite[Theorem~4]{blanco2018rank}.
   	
\begin{corollary}
\label{cor:boundR1}
	Let $\alpha$ be the negative of the sum of the reciprocal roots of the $j$-th generalized tensor zeta polynomial $P_\C^{(j)}(T)$. If $t_j^\R< n$ then 
	\begin{equation*}
		t_j^\R\leq \frac{n}{n_1}\log_q\left(\left(\alpha+\bbin{j+1}{1}{\frac{n}{n_1}}\right)(q-1)+1\right)-\frac{n}{n_1}.
	\end{equation*}
\end{corollary}
\begin{proof}
	For ease of notation, throughout this proof we write $b_a^{(j)}$ and $p_a^{(j)}$ instead of $b_a^{(\R,j)}(\C)$ and $p_a^{(\R,j)}(\C)$. As a consequence of Theorem~\ref{thm:zetapol}, we have
	\begin{equation*}
		b_0^{(j)}=p_0^{(j)}\qquad \textup{ and }\qquad b_{\frac{n}{n_1}}^{(j)}=p_\frac{n}{n_1}^{(j)}+\bbin{j+1}{1}{\frac{n}{n_1}}b_0^{(j)}=\alpha\, p_0^{(j)}+\bbin{j+1}{1}{\frac{n}{n_1}}p_0^{(j)}
	\end{equation*}
	where the last equality follows from the fact that $P_\C(T)^{(j)}=p_0^{(j)}(1+\alpha T+\cdots)$. In particular, we have
	\begin{equation}
		\label{eq2:boundR}
		\frac{b_{\frac{n}{n_1}}^{(j)}}{b_0^{(j)}}=\frac{\alpha\, p_0^{(j)}+\bbin{j+1}{1}{\frac{n}{n_1}}p_0^{(j)}}{p_0^{(j)}}=\alpha+\bbin{j+1}{1}{\frac{n}{n_1}}.
	\end{equation}
	The statement follows by substituting~\eqref{eq2:boundR} into the bound of Theorem~\ref{thm:boundR}.
\end{proof}

The following result extends Proposition~\ref{prop:proptR}.

\begin{proposition}
	The following hold.
	\begin{enumerate}[label=(\arabic*)]
	\setlength\itemsep{0.5em}
		\item $t_j\geq j$.
		\item $t_j^\D\leq\min\left\{n-\frac{n}{n_i}\left\lfloor\frac{n_i}{n}(k-j)\right\rfloor:i\in[r]\right\}$.
	\end{enumerate}
\end{proposition}
\begin{proof}
	Clearly, if $A\in\A$ is such that $\dimq{A\cap\C}\geq j$ then $\dimq{A}\geq j$ which implies the first part of the statement.
	It is not hard to check that an argument similar to the one in the proof of Proposition~\ref{prop:proptR}\eqref{item1:proptR} implies
	\begin{equation*}
		\min\left\{\[\dd(A)\]:A\in\A^\D\mid \dd(A)\in\mfN_i^\D \textup{ and } \dd(A\cap\C)\geq j \right\}\leq n-\frac{n}{n_i}\left\lfloor\frac{n_i}{n}(k-j)\right\rfloor
	\end{equation*}
	for all $i\in[r]$. Therefore, by Definition~\ref{def:tj} we get
	\begin{align*}
		t_j^\D=&\min\{\[\dd(A)\]:A\in\A^\D\mid \dd(A\cap\C)\geq j\}\\
		&=\min\bigcup_{i=1}^r\left\{\[\dd(A)\]:A\in\A^\D\mid \dd(A)\in\mfN_i^\D \textup{ and } \dd(A\cap\C)\geq j \right\}\\
		&\leq\min\left\{n-\frac{n}{n_i}\left\lfloor\frac{n_i}{n}(k-j)\right\rfloor:i\in[r]\right\}.
	\end{align*}
	This concludes the proof.
\end{proof}

We conclude this section by deriving bounds for the tensor weight related to $\A^\D$ which is similar to the one in Theorem~\ref{thm:boundR}.

\begin{theorem}
\label{thm:boundD}
	For all $i\in[r]$, we let $\e^{(i)}$ be the element of $\mathfrak{Z}^\cl$ with $1$ in position $i$ and zeros elsewhere and $\c^{(j,i)}$ to be the minimal element $\a$ of $\mfN_i^\D$ such that $\[\a\]\geq t_j^\D$, that is
	$$\c^{(j,i)}:=\bigwedge\{\a\in\mfN_i^\D:\[\a\]\geq t_j^\D\}.$$
	We have
	\begin{equation*}
		t_j^\D\leq\min\left\{\frac{n}{n_i}\log_q\left(\frac{b_{\c^{(j,i)}+\e^{(i)}}}{b_{\c^{(j,i)}}}(q-1)+1\right)-\frac{n}{n_i}:i \in [r]\mid \c^{(j,i)}<\n\right\}.
	\end{equation*}
\end{theorem}
\begin{proof}
	For the ease of notation, throughout this proof we write $W_\b^{(j)}$ and $B_\b^{(j)}$ instead of $W_\b^{(\D,j)}$ and $B_\b^{(\D,j)}$ for all $\b\in\mfN^\D\cup\{\o\}$, and $b_\a^{(j)}$ and $p_\a^{(j)}$ instead of $b_\a^{(j)}(\C)$ and $p_\a^{(j)}(\C)$, for all $\a\in\mathfrak{Z}^\D$.  It is sufficient to show that
	\begin{equation*}
		t_j^\D\leq\frac{n}{n_i}\log_q\left(\frac{b_{\c^{(j,i)}+\e^{(i)}}}{b_{\c^{(j,i)}}}(q-1)+1\right)-\frac{n}{n_i}.
	\end{equation*}
	for an $i\in[r]$ such that $n_i\in\Delta$ and $\c^{(j,i)}< \n$, since $\mfN_i^\R\cap\mfN_s^\R=\{\n\}$ for all $i,s\in[r]$. Fix such $i\in [r]$ and observe that 
	\begin{equation*}
		\left\{\c^{(j,i)}+\e^{(i)}-s\cdot\e^{(i)}\in\mfN^\D:s \in[j]\cup\{0\}\right\}=\left\{\c^{(j,i)},\c^{(j,i)}+\e^{(i)}\right\}
	\end{equation*}
	and that $B_\a^{(j)}=0$ for all $\a\in\mfN_i^\D$ such that $\[\a\]<t_j^\D$. Therefore, we get
	\begin{align*}
		W_{\c^{(j,i)}+\e^{(i)}}^{(j)}&=B_{\c^{(j,i)}+\e^{(i)}}^{(j)}-B_{\c^{(j,i)}}^{(j)}\qbin{n_i-\c^{(j,i)}_i}{1}\\
		&=b_{\c^{(j,i)}+\e^{(i)}}^{(j)}\qbin{n_i}{\c^{(j,i)}_i+1}-b_{\c^{(j,i)}}^{(j)}\qbin{n_i}{\c^{(j,i)}}\qbin{n_i-\c^{(j,i)}_i}{1}
	\end{align*}
	where the last equality follows by Definition~\ref{def:norbcl}. Observe that $W_{\c^{(j,i)}+\e^{(i)}}^{(j)}\geq 0$ which implies
	\begin{align*}
		0&\leq b_{\c^{(j,i)}+\e^{(i)}}^{(j)}\qbin{n_i}{\c^{(j,i)}_i+1}-b_{\c^{(j,i)}}^{(j)}\qbin{n_i}{\c^{(j,i)}_i}\qbin{n_i-\c^{(j,i)}_i}{1}\\
		&=b_{\c^{(j,i)}+\e^{(i)}}^{(j)}\frac{q^{n_i-\c^{(j,i)}_i}-1}{q^{\c^{(j,i)}_i-1}-1}\qbin{n_i}{\c^{(j,i)}_i}-b_{\c^{(j,i)}}^{(j)}\qbin{n_i}{\c^{(j,i)}_i}\qbin{n_i-\c^{(j,i)}_i}{1}.
	\end{align*}
	Since $\qbin{n_i}{\c^{(j,i)}_i}> 0$ and $b_{\c^{(j,i)}}^{(j)}>0$, we get
	\begin{equation*}
		0\leq \frac{b_{\c^{(j,i)}+\e^{(i)}}}{b_{\c^{(j,i)}}}\cdot\frac{q^{n_i-\c^{(j,i)}_i}-1}{q^{\c^{(j,i)}_i-1}-1}-\qbin{n_i-\c^{(j,i)}_i}{1}=\frac{b_{\c^{(j,i)}+\e^{(i)}}}{b_{\c^{(j,i)}}}\cdot\frac{q^{n_i-\c^{(j,i)}_i}-1}{q^{\c^{(j,i)}_i-1}-1}-\frac{q^{n_i-\c^{(j,i)}_i}-1}{q-1}
	\end{equation*}
	which implies
	\begin{equation*}
		q^{\c^{(j,i)}_i+1}-1\leq \frac{b_{\c^{(j,i)}+\e^{(i)}}}{b_{\c^{(j,i)}}}(q-1)
	\end{equation*}
	and therefore the statement.
\end{proof}

As a consequence we have the following.

\begin{corollary}
\label{cor:boundR}
	For all $i\in[r]$, we let $\e^{(i)}$ be the element of $\mathfrak{Z}^\cl$ with $1$ in position $i$ and zeros elsewhere and $\c^{(j,i)}$ to be the minimal element $\a$ of $\mfN_i^\R$ such that $\[\a\]\geq t_j^\R$, that is
	$$\c^{(j,i)}:=\bigwedge\{\a\in\mfN_i^\R:\[\a\]\geq t_j^\R\}.$$
	We have
	\begin{align*}
		t_j^\R &\leq \min\left\{\frac{n}{n_1}\log_q\left(\frac{b_{\c^{(j,i)}+\e^{(i)}}}{b_{\c^{(j,i)}}}(q-1)+1\right)-\frac{n}{n_1}:i \in \delta_1\mid \c^{(j,i)}<\n\right\}.
	\end{align*}
\end{corollary}

\begin{remark}
	It is not hard to check that if $|\delta_1|=1$, i.e. if 
	\begin{equation*}
		\A^\R=\left\{A^{(1)}\otimes\bigotimes_{i=2}^r\Fq^{n_i}:A^{(1)}\leq\Fq^{n_1}\right\},
	\end{equation*} 
	then the bounds in Corollary~\ref{cor:boundR} reduce to the ones in Theorem~\ref{thm:boundR} and Corollary~\ref{cor:boundR1}.
\end{remark}

\section*{Acknowledgements}

The author is very grateful to Eimear Byrne for fruitful discussions on this topic and for help in improving Theorem~\ref{thm:zetapolRrefined}.

\bibliographystyle{abbrv} 
\bibliography{zfbiblio.bib}  
\end{document}